\documentclass[preprint,showpacs,preprintnumbers,amsmath,amssymb] {revtex4}
\usepackage{graphicx}
\usepackage{dcolumn}
\usepackage{bm}
\newcommand{\bee}{\begin{eqnarray}}
\newcommand{\eend}{\end{eqnarray}}
\newcommand{\rmd}{{\rm d}}
\newcommand{\rme}{{\rm e}}
\newcommand{\rmi}{{\rm i}}
\newcommand{\bea}{\begin{eqnarray}}
\newcommand{\eea}{\end{eqnarray}}

\begin{document}

\title{Convexity of effective Lagrangian in nonlinear electrodynamics as derived from causality}

\author{Anatoly E. Shabad$^1$ and Vladimir V. Usov$^2$}
\affiliation{ $^1$ P.N. Lebedev Physics Institute, Moscow 117924,
Russia\\
$^2$Center for Astrophysics, Weizmann Institute, Rehovot 76100,
Israel}
\begin{abstract} In nonlinear electrodynamics, by implementing the causality
 principle as the requirement  that the group velocity of elementary excitations
 over a background field should not exceed unity, and the unitarity principle
 as the requirement that the residue of the propagator should be
 nonnegative, we find restrictions on the behavior of massive and
 massless dispersion curves and establish the convexity of the
 effective Lagrangian on the class of constant fields, also
 the positivity of all characteristic dielectric and magnetic permittivity constants.
 Violation of the general principles by the one-loop approximation in QED at exponentially large magnetic
field is analyzed  resulting in complex energy tachyons and
super-luminal ghosts that signal the instability of the magnetized
vacuum. General grounds for kinematical selection rules in the
process of photon splitting/merging are discussed.
\end{abstract}
\maketitle
\section{Introduction} The effective action that is
defined as the Legendre transform of the generating functional of
the Green functions \cite{weinberg}  and, in its turn, is itself a
generating functional of the (one-particle-irreducible) vertices
makes a basic quantity in  quantum field theory. This is a
c-numerical functional of fields and their derivatives, a knowledge
of which is meant to supply one with the final solution to the
theory. For this reason it seems important to see, how the most
fundamental principles manifest themselves as some general
properties of the effective action to be respected by model- or
approximation-dependent calculations, and whose violation might
signal important inconsistencies in the theory underlying these
calculations. Such inconsistencies may show themselves first of all
as ghosts and tachyons, that play an important role \cite{aref'eva}
in cosmological speculations about  forming the $\Lambda$-term and
dark energy using a scalar (Higgs) field yet to be discovered in the
coming experiments on the Large Hadronic Collider.

It is stated \cite{weinberg} basing on a formal continual integral
representation for the propagator that, when the effective action
$\Gamma(\phi)$ of a scalar field with mass $m$ is considered, its
second variational derivative
$\Sigma(x-y|\phi_0)=\left.\delta^2\Gamma/\delta\phi
(x)\delta\phi(y)\right|_{\phi=\phi_0}$
calculated at the constant background value of this field,
$\phi(x)=\phi_0,$ i.e. the  mass operator against this background,
is a nonpositive quantity, $\Sigma\leq 0$. In other words, the
effective Lagrangian is expected -- to the extent that that formal
property survives  perturbative or other calculations -- to be a
nonconvex (while the effective potential to be a convex) function of
a constant scalar field. However, the same statement may be
considered as the one directly prescribed by  the causality
principle. Indeed, the spectral curve of small excitations over the
constant field background, $k_0=\sqrt{{\bf k}^2+m^2-\Sigma (k)},$
where $k=(k_0,\bf k)$ is the (4-momentum) variable,
Fourier-conjugate to the 4-coordinate difference $x-y$, satisfies
the causal propagation condition reading that its group velocity
should not exceed unity, the absolute speed limit for any signal,
$|\partial k_0/\partial {\bf k}|=|\bf k|/k_0 \leq 1$ for any
nonnegative  mass squared $m^2\geq 0$, provided, again, that
$\Sigma\leq 0$.

The case under our consideration here is much less trivial as we
deal not with a massive scalar, but with a massless vector gauge
field. The results apply, first of all, to electromagnetic field,
but also -- in a restrictive way -- to nonabelian gluon fields.
Nonlinear electrodynamical models are also considered for
cosmological purposes \cite{novello-1} with the advantage that
instead of the scalar field, uncertain to be physically identified,
only well established electromagnetic field is involved.

We are going to demonstrate that the requirement of the causal
propagation of elementary excitations over the vacuum occupied by a
background field with a constant and homogeneous field strength,
supplemented by the requirements of translation-, Lorentz-, gauge-,
P- and C- invariances and unitarity has a direct impact on the
effective Lagrangian. For  the case - which is general for
electromagnetic field, but special for a nonabelian field - where
the Lagrangian depends on gauge-invariant combinations (field
strengthes) $F_{\alpha\beta}(z)=\partial_\alpha
A_\beta(z)-\partial_\beta A_\alpha(z)$ of the background field
potentials we make sure that the above requirements are expressed as
certain inequalities to be obeyed by the effective Lagrangian and
its first and second derivatives with respect to the two {\em field
invariants} $\mathfrak{F}=\frac
1{4}F_{\rho\sigma}F_{\rho\sigma}=\frac 1{2}(B^2-E^2)$ and
$\mathfrak{G}=\frac
1{4}F_{\rho\sigma}\tilde{F}_{\rho\sigma}=(\bf{EB}),$ where $\bf E$
and $\bf B$ are background electric and magnetic fields,
respectively, and the dual field tensor is defined as
$\tilde{F}_{\rho\sigma}=\frac
1{2}\epsilon_{\rho\sigma\lambda\kappa}F_{\lambda\kappa}$, where the
completely antisymmetric unit tensor is defined in such a way that
$\epsilon_{1230}=1$. More specifically, we demonstrate that it is a
convex function with respect to the both variables $\mathfrak{F,G}$
for any constant value of $\mathfrak{F}\gtrless 0$ and
$\mathfrak{G}=0$.

In Section II  model- and approximation-independent study is
undertaken.

In Subsection A we are basing on the general diagonal representation
of the polarization operator and photon Green function in terms of
its eigenvectors and eigenvalues, obtained for arbitrary values of
the momentum $k$ and for nonzero constant field invariants
$\mathfrak{F,G}$ in \cite{batalin}, to find limitations on the
location of dispersion curves, imposed  by  demanding that the group
velocity of the vacuum excitations be less than/or equal to unity.
We find that the massless  branches of these curves ("photons"),
whose existence is always guarantied by the gauge invariance, for
every polarization mode are outside  the light cone (or on it) in
the momentum space $k^2=0$, whereas the massive branches all should
pass below a certain curve in the plane $(k_0^2-k_3^2, k_\perp^2)$,
where $k_3$ and $\bf k_\perp$ are the excitation momentum components
along and across the direction of the background magnetic and
electric fields in the special frame, where these are mutually
parallel.

In Subsection B we confine ourselves to the infrared asymptotic
behavior $k_\mu\rightarrow0$ of the polarization operator, in which
case its eigenvalues can be expressed in terms of first and second
derivatives of the effective Lagrangian with respect to the field
invariants $\mathfrak{F,G}$. Massless dispersion curves are
explicitly found in terms of these derivatives for the
"magnetic-like" case $\mathfrak{F}>0$, $\mathfrak{G}=0.$ The
restrictions of Subsection A, now supplemented with the unitarity
requirement that the residue of the Green function in the pole,
corresponding to
 the mass shell of the elementary excitation, be nonnegative (completeness of
the set of states with nonnegative norm),  actualize as a number of
inequalities, to be satisfied by these derivatives. They mean, in
particular, that the effective Lagrangian is a convex function of
the field invariants in the point  $\mathfrak{G}=0.$ Basing on the
study made in Appendix we reveal the physical sense of the
quantities subject to these inequalities as dielectric and magnetic
permeabilities responsible for polarizing small static charges and
currents of special configurations (There is no universal linear
response function able to cover every configuration). In Subsection
C the inequalities of Subsection B are extended to include also the
"electric-like" background field $\mathfrak{F}<0$, $\mathfrak{G}=0,$
so in the end the whole axis of the variable $\mathfrak{F}$ is
included into result.

In Subsection D we find the contribution of the polarization
operator into effective Lagrangian, which is local in the infrared
limit and presents the Lagrangian for small, slow, long-wave
perturbations of the background field. This gives us the possiblity
to define their energy-momentum tensor via the Noether theorem. By
imposing the conditions of the positivity of the energy density and
of non-spacelikeness of the energy-momentum flux vector -- the Weak
Energy Condition  and Dominant Energy Condition of Hawking and Ellis
-- that might be considered as an alternative to the requirements
used in Subsections A and B, we find some inequalities that do not
contradict to those found in the previous subsection, but are
weaker. This urges us to make an important admission  that those
popular conditions may be in a certain respect insufficient.

In Section III we test the restrictions obtained in Section II for
the Euler-Heisenberg one-loop effective Lagrangian of Quantum
Electrodynamics and for the Lagrangian of Born-Infeld to establish
that the latter perfectly satisfies all of them. On the contrary,
some of them are violated by the Euler-Heisenberg Lagrangian at
exponentially large magnetic field, leading to appearance of ghosts
and tachyons, signifying the instability of the magnetized vacuum
due to the lack of asymptotic freedom in QED. (The instability of
the electrified vacuum in this approximation thanks to Schwinger's
electron-positron pair creation goes without saying). It is a
surprise that the convexity itself is not violated at any value of
the magnetic field.

In Section IV we decided to use the  opportunity, presented by the
fact that all the appropriate circumstances have been exposed, to
discuss a somewhat different matter about S.Adler's kinematical
selection rule that is established by appealing to one-loop
approximation and forbids some transitions between photon modes in
the cubic process of the photon splitting in a strong magnetic
field, what is important for  formation of radiation in the pulsar
magnetosphere. Within our context this rule is an inequality between
  derivatives of the effective Lagrangian,  involved in the
previous analysis. We propose arguments that may rule out a
violation of Adler's kinematical selection rule basing on dual
symmetry consideration.

In concluding Section V we perform an attempt of comparative
discussion of our approach with other  ways of introducing causality
into consideration.

\section{Unitarity and causality}
\subsection{Configuration of exact dispersion curves}
Let
$\mathfrak{L}(z
)$ be the nonlinear part of the effective Lagrangian as a function
of the two electromagnetic field invariants $\mathfrak{F}$ and
$\mathfrak{G}$ and, generally, of other Lorentz scalars that can be
formed by the electromagnetic field tensor $F_{\mu\nu}$ and its
space-time derivatives.  The total action  is $S_{\rm tot}=\int
L_{\rm tot}(z)\rmd^4z$, where $ L_{\rm tot}(z)=-\mathfrak{F}(z)+
\mathfrak{L}(z).$ 
It is assumed that \bee\label{corresp}\left.\frac{\delta
\Gamma}{\delta\mathfrak{F}}\right|_{\mathfrak{F}=\mathfrak{G}=0}=0,\eend
where $\Gamma=\int \mathfrak{L}(z)\rmd^4z$, according to the
correspondence principle, since $-\mathfrak{F}$ is the classical
Lagrangian.

We consider the  background field, which is constant in time and
space and has only one nonvanishing invariant: $\mathfrak{F}\neq 0,
\mathfrak{G}=0$ (although $\mathfrak{G}$
 may be involved in  intermediate equations).
 This field is purely magnetic in
a special Lorentz frame, if $\mathfrak{F}>0$, and purely electric in
the opposite case, $\mathfrak{F}<0$. Such fields will be called
magnetic- or electric-like, respectively. 

 Polarization operator is responsible for  small perturbations
 above the constant-field background. In accordance with the role
 of the effective action as the generating functional of vertex functions,
 the polarization operator  is defined as the
second variational derivative with respect to the vector potentials
$A_\mu$\bee\label{Pi} \Pi_{\mu\tau}(x,y)=
\left.\frac{\delta^2S}{\delta A_\mu(x)\delta A_\tau(y)}\right
|_{\mathfrak{G}=0, \mathfrak{F}= \rm const}.\eend The action $S$
here is meant to be - prior to the two differentiations over $A_\mu,
A_\tau$ -  a functional containing field derivatives of arbitrary
order, but the fields are set constant after the differentiations.
Nevertheless, their derivatives do contribute into the polarization
operator (\ref{Pi}) leading to its complicated dependence on the
momentum $k$, the variable, Fourier conjugated to $(x-y)$.

It follows from the translation- Lorentz-, gauge-, P- and
charge-invariance \cite{batalin, annphys,shabtrudy} that the Fourier
transform of the tensor (\ref{Pi}) is diagonal \bee\label{pimunu}
\Pi_{\mu\tau}(k,p)=\delta(k-p)\Pi_{\mu\tau}(k),\qquad
\Pi_{\mu\tau}(k)=\sum_{a=1}^3\kappa_a(k)~\frac{\flat_\mu^{(a)}~
\flat_\tau^{(a)}}{(\flat^{(a)})^2}\eend  in the following
basis:\begin{eqnarray}\label{vectors} \flat^{(1)}_\mu=(F^2k)_\mu
k^2-k_\mu(kF^2k),\quad \flat_\mu^{(2)}=(\tilde{F}k)_\mu,\quad
\flat_\mu^{(3)}=(Fk)_\mu,\quad \flat_\mu^{(4)}=k_\mu,
\end{eqnarray} where $(\tilde{F}k)_\mu\equiv \tilde{F}_{\mu\tau}k_\tau$, $(Fk)_\mu\equiv
F_{\mu\tau}k_\tau$, $(F^2k)_\mu\equiv F^2_{\mu\tau}k_\tau$,
$kF^2k\equiv k_\mu F^2_{\mu\tau}k_\tau$, formed by the eigenvectors
of the polarization operator 
\begin{eqnarray}\label{eigen}\Pi_{\mu\tau}~\flat^{(a)}_\tau=\kappa_a(k)~\flat^{(a)}_\mu.
\end{eqnarray} We are working in Euclidian metrics with the results
analytically continued to Minkowsky space, hence we do not
distinguish between co- and contravariant indices. All eigenvectors
are mutually orthogonal, $\flat^{(a)}_\mu \flat_\mu^{(b)}\sim
\delta_{ab}$, this means that the first three ones are
4-transversal, $\flat^{(a)}_\mu k_\mu=0$; correspondingly
$\kappa_4=0$ as a consequence of the 4-transversality of the
polarization operator. The unit matrix 
is decomposed as \bee\label{unit}
\delta_{\mu\tau}=\sum_{a=1}^4\frac{\flat_\mu^{(a)}~
\flat_\tau^{(a)}}{(\flat^{(a)})^2}\qquad {\rm or}\qquad
\delta_{\mu\tau}-\frac{k_\mu k_\tau}{k^2}
=\sum_{a=1}^3\frac{\flat_\mu^{(a)}~
\flat_\tau^{(a)}}{(\flat^{(a)})^2}.\eend The eigenvalues
$\kappa_a(k)$ of the polarization operator are scalars and depend on
$\mathfrak{F}$ and on any two of the three momentum-containing
Lorentz invariants $k^2={\bf k}^2-k_0^2,\; kF^2k,\;k\tilde{F}^2k$ ,
subject to one relation $\frac{k\tilde{F}^2k}{2\mathfrak{F}}-k^2=
\frac{k{F}^2k}{2\mathfrak{F}}$.
 The squares of the eigenvectors
are\bee\label{squares}
(\flat^{(1)})^2=-(kF^2k)((kF^2k)+2\mathfrak{F}k^2)=k^2k_\perp^2(2\mathfrak{F})^2(k_3^2-k^2_0),\nonumber\\
\quad (\flat^{(2)})^2=-(kF^2k),\quad
(\flat^{(3)})^2=-(k\widetilde{F}^2k)\quad\eend

The diagonal representation of the photon Green function as an exact
solution to the Schwinger-Dyson equation with the polarization
operator (\ref{pimunu}) taken for the kernel is (up to arbitrary
longitudinal part):
\begin{eqnarray}\label{photon2} D_{\mu\tau}(k)=
\sum_{a=1}^4 D_a(k)~\frac{\flat_\mu^{(a)}~
\flat_\tau^{(a)}}{(\flat^{(a)})^2},\nonumber\\
D_a(k)=\left\{\begin{tabular}{cc}$(k^2-\kappa_a(k))^{-1},$&\qquad\;
$a=1,2,3$\\arbitrary, &~ $a$ = 4\end{tabular}\;.\right.
\end{eqnarray}
The dispersion equations that define the mass shells of the three
eigen-modes are
\bee\label{dispersion}\kappa_a(k^2,\frac{k{F}^2k}{2\mathfrak{F}},\mathfrak{F})=k^2,\qquad
a=1,2,3. \eend

All the equations above are valid both for magnetic- and
electric-like cases, $\mathfrak{F}\lessgtr 0$, $\mathfrak{G}=0$. 
If, specifically,  the magnetic-like background field $\mathfrak{F}>
0$, $\mathfrak{G}=0$ is considered,  in the special frame the
field-containing invariants become \bee\label{special}\frac{k\tilde
{F}^2k}{2\mathfrak{F}}=k_3^2-k_0^2,\qquad\frac{k{F}^2k}{2\mathfrak{F}}=-k_\perp^2,\qquad
\mathfrak{F}=\frac{B^2}2,\eend where we directed the magnetic field
$\bf B$ along the axis 3, and the two-dimensional vector ${\bf
k}_\perp$ is the photon momentum projection onto the plane
orthogonal to it. On the contrary, if we deal with the electric-like
background field $\mathfrak{F}< 0$, $\mathfrak{G}=0,$ in the special
frame, where only electric field $\bf E$ exists and is
 directed along axis 3, we have, instead of (\ref{special}), the following
  relations for the background-field- and momentum-containing invariants
\bee\label{special2}\frac{k\tilde
{F}^2k}{2\mathfrak{F}}=k_\perp^2,\qquad\frac{k{F}^2k}{2\mathfrak{F}}=k_0^2-k_3^2,\qquad
\mathfrak{F}=\frac{-E^2}2,\eend where  the two-dimensional vector
${\bf k}_\perp$ now is the photon momentum projection onto the plane
orthogonal to $\bf E$. In the both cases the dispersion equations
(\ref{dispersion}) can be represented in the same form
\bee\label{dispersion2}\kappa_a(k^2,k_\perp^2,\mathfrak{F})=k^2,\qquad
a=1,2,3 \eend and their solutions have the following general
structure, provided by relativistic invariance \bea
k_0^2=k_3^2+f_a(k_\perp^2),\hspace{5mm} a=1,2,3\label{law}. \eea

It is notable that the structure (\ref{law}) retains when the second
invariant is also nonzero, $\mathfrak{G}\neq 0,$ this time the
direction 3 being the common direction of the background electric
and magnetic fields in the special reference frame, where these are
mutually parallel. Hence, the restrictions on the way the dispersion
curves pass to be obtained below in the present subsection will
remain valid in this general case, too. The only specific feature of
the general case is that the eigenvectors $\flat_\mu$ are no longer
given by the final expressions (\ref{vectors}), but are now  linear
combinations of the vectors (\ref{vectors}) with generally unknown
coefficients depending on the scalar combinations of the background
field and momentum \cite{batalin}, \cite{shabtrudy}.

The causality principle requires that the modulus of the group
velocity, calculated on each mass shell (\ref{law}), be less or
equal to the speed of light in the free vacuum $c=1$:
\bee\label{group}|\textbf{v}_{\rm gr}|^2=\left(\frac{\partial
k_0}{\partial k_3}\right)^2+\left|\frac{\partial k_0}{\partial
\textbf{k}_\perp}\right|^2=\frac{k_3^2}{k_0^2}+\left|\frac{\textbf{k}_\perp}{k_0}\cdot
f_a^\prime\right|^2
=\frac{k_3^2+k_\perp^2\cdot(f_a^\prime)^2}{k_3^2+f_a(k_\perp^2)}\leq
1,\eend where $f_a^\prime=\rmd f_a(k_\perp^2)/\rmd k_\perp^2$. This
imposes the obligatory condition on the form and location of the
dispersion curves (\ref{law}), i.e. on the function
$f_a(k_\perp^2)$, to be fulfilled within every reasonable
approximation (remind that $k_3^2+f_a(k_\perp^2)\geq 0 $ due to
(\ref{law})) :\bee\label{causality} k_\perp^2\left(\frac{\rmd
f_a(k_\perp^2)}{\rmd k_\perp^2}\right)^2\leq f_a(k_\perp^2).\eend
This inequality requires first of all that $f_a(k_\perp^2)\geq 0$,
hence no branch of any dispersion curve may get into the region
$k_0^2-k_3^2<0$. If it might, the photon energy $k_0$ would have an
imaginary part within the momentum interval
$0<k_3^2<-f_a(k_\perp^2),$ corresponding to the vacuum excitation
exponentially growing in time. This sort of tachyon would signal the
instability of the magnetized vacuum. Inequality (\ref{causality})
further requires that \bee\label{rest}\frac{\rmd
f_a^{\frac1{2}}(k_\perp^2)}{\rmd k_\perp}\leq 1,\qquad {\rm
or}\qquad f_a^{\frac1{2}}(k_\perp^2)\leq const+k_\perp.\eend

The unitarity imposes the limitations that the residues of the
photon propagator (\ref{photon2}) in the poles corresponding to
every photon mass shell (\ref{dispersion}) be nonnegative - the
positive definiteness of the norm of every elementary excitation of
the vacuum. This requirement implies:\bee\label{unitarity}
1-\left.\frac{\partial\kappa_a(k^2,k_\perp^2,\mathfrak{F})}{\partial
k^2}\right|_{k_0^2-k_3^2=f_a(k_\perp^2)}\geq 0.\eend

We shall prove somewhat later (see eq. (\ref{kappa}) below) that
\bea \kappa_a(0,0,\mathfrak{F})=0, \hspace{5mm}
a=1,2,3.\label{gauge} \eea This property implies that for each mode
there always exists a dispersion curve with $f_a(0)=0$, which passes
through the origin in the $(k_0^2-k_\parallel^2, k_\perp^2)$-plane.
It is such branches that are called photons, since they are massless
in the sense that the energy $k_0$ turns to zero for the particle at
rest, $k_3=k_\perp=0$ (although, generally, $k^2\neq 0$ where $\bf
k\neq 0$). Other branches for each mode may also appear provided
that a dynamical model includes an existence of a massive excitation
of the vacuum with quantum numbers of a photon, for instance the
positronium atom \cite{ass} or a massive axion. Thus, \emph{for
photons}, the integration constant $const$ in (\ref{rest}) should be
chosen as zero. We conclude that the causality requires that in the
plane $(\sqrt{k_0^2-k_3^2}, k_\perp)$ the photon dispersion curves
are located outside the light cone: $k^2 \geq 0$. (Remind that the
light cone $k^2= 0$  is the mass shell of a photon in the vacuum
without an external field.) However, unlike the previous case, a
violation of this ban would not lead to a complex-energy tachyon or
directly signalize the vacuum instability.

The refraction index squared $n^2_a$  is defined for photons of mode
\emph{a} on the mass shell (\ref{law}) as \bea\label{refrindex}
n_a^2\equiv\frac{|{\bf
k}|^2}{k_0^2}
=1+\frac{k_\bot^2-f_a(k_\bot^2)}{k_0^2} .\eend It follows from
 (\ref{rest}) with $const=0$ that the refraction index is greater
 than unity - the statement common in standard optics of media
 (this is not, certainly, true for (massive) plasmon branches).
Consequently, the modulus of the phase velocity in each mode ${\bf
v}^{\rm ph}_a=\frac {k_0}{|\bf k|}\frac{\bf k}{|\bf k|}$ equal to
$1/n_a$ is, {\em for the  photon proper}, also smaller than the
velocity of light in the vacuum $c=1$. This is not the case for a
massive -- {\em e.g.} positronium -- branch of the photon
dispersion curve, where ${|\bf v}^{\rm ph}_a|>1$ without any
importance for causality.

Now that we established that for photons one has ${\bf k}^2\geq
k_0^2$, or $k^2\geq 0$, we see from the dispersion equation
(\ref{dispersion})  that the eigenvalues $\kappa_a$ are nonnegative
in the momentum region, where the photon dispersion curves lie, i.e.
the polarization operator is nonnegatively defined matrix there.
\subsection{Infrared limit: properties of the Lagrangian as a
function of constant fields} Hitherto, we were dealing with the
elementary excitation of arbitrary 4-momentum $k_\mu.$ To get the
(infrared) behavior of the polarization operator at $k_\mu\sim 0$ it
is sufficient to have at one's disposal the effective Lagrangian as
a function of constant field strengthes, since their derivatives, if
included in the Lagrangian, would supply extra powers of the
momentum $k$ in the expression (\ref{Pi}) for the polarization
operator. We shall restrict ourselves to the infrared asymptotic
below. Our goal is to establish some inequalities imposed on the
derivatives of the effective Lagrangian $\mathfrak{L}$ over the
constant fields by the requirement (\ref{causality}) that any
elementary excitation of the vacuum should not propagate with the
group velocity larger than unity and the requirement
(\ref{unitarity}) that the residue of the Green function is positive
in the photon pole. To proceed beyond this limit we had to include
the space and time derivatives of the fields into the Lagrangian.
Then, utilizing the same  requirements (\ref{causality}),
(\ref{unitarity}) the results concerning the convexity of the
effective Lagrangian with respect to the constant fields to be
obtained below, might be, perhaps, extended to convexities with
respect to the derivative-containing field variables.

Aiming at the infrared limit we do not include time- and
space-derivatives of the field strengthes in the equations that
follow. Using the definition $F_{\alpha\beta}(z)=\partial_\alpha
A_\beta(z)-\partial_\beta A_\alpha(z)$ we find
\bee\label{fi}\frac\delta {\delta A_\mu(x)}\int
\mathfrak{F}(z)\rmd^4z=\int F_{\alpha\mu}(z)\frac{\partial}{\partial
z_\alpha}\delta^4(x-z)\rmd^4z,\nonumber\\ \frac\delta {\delta
A_\mu(x)}\int \mathfrak{G}(z)\rmd^4z=\int
\tilde{F}_{\alpha\mu}(z)\frac{\partial}{\partial
z_\alpha}\delta^4(x-z)\rmd^4z.\eend Then, for the first variational
derivative of the action one has\bee\label{firstder}\frac{\delta
\Gamma} {\delta A_\mu(x)}=
\int\left[\frac{\partial\mathfrak{L}(\mathfrak{F}(z),\mathfrak{G}(z))}{\partial
\mathfrak{F}(z)}F_{\alpha\mu}(z)+\frac{\partial\mathfrak{L}(\mathfrak{F}(z),\mathfrak{G}(z))}{\partial
\mathfrak{G}(z)}\tilde{F}_{\alpha\mu}(z)\right]\frac{\partial}{\partial
z_\alpha}\delta^4(x-z)\rmd^4z.\eend By repeatedly applying eq.
(\ref{firstder})  we get for the infrared (IR) limit of the
polarization operator in a constant external field
\bee\label{secdir}\Pi^{\rm
IR}_{\mu\tau}(x,y)=\left.\frac{\delta^2\Gamma}{\delta A_\mu(x)\delta
A_\tau(y)}\right |_{\mathfrak {F},\mathfrak {G} =\rm const}=
\left\{\frac{\partial\mathfrak{L}(\mathfrak{F}(z),\mathfrak{G}(z))}{\partial
\mathfrak{F}(z)}\left(\frac{\partial^2}{\partial x_\tau\partial
x_\mu}-\Box\delta_{\mu\tau}\right)\right.
-\nonumber\\\nonumber\\-\left.\frac{\partial^2\mathfrak{L}(\mathfrak{F}(z),\mathfrak{G}(z))}{\partial
(\mathfrak{F}(z))^2}\left(F_{\alpha\mu}\frac{\partial }{\partial
x_\alpha}\right)\left(F_{\beta\tau}\frac{\partial }{\partial
x_\beta}\right)-
\frac{\partial^2\mathfrak{L}(\mathfrak{F}(z),\mathfrak{G}(z))}{\partial
(\mathfrak{G}(z))^2}\left(\tilde{F}_{\alpha\mu}\frac{\partial
}{\partial x_\alpha}\right)\left(\tilde{F}_{\beta\tau}\frac{\partial
}{\partial x_\beta}\right)\right.
-\nonumber\\
\nonumber\\
-\left.\frac{\partial^2\mathfrak{L}(\mathfrak{F}(z),\mathfrak{G}(z))}{\partial
\mathfrak{F}(z)\partial \mathfrak{G}(z)}\left[
\left({F}_{\alpha\mu}\frac{\partial }{\partial
x_\alpha}\right)\left(\tilde{F}_{\beta\tau}\frac{\partial }{\partial
x_\beta}\right)+\left(\tilde{F}_{\alpha\mu}\frac{\partial }{\partial
x_\alpha}\right)\left({F}_{\beta\tau}\frac{\partial }{\partial
x_\beta}\right) \right]\right\}_{F=\rm
const}\delta^4(x-y).\quad\eend The P-invariance requires that the
effective Lagrangian should be an even function of the pseudoscalar
$\mathfrak{G}$. Hence all the terms in the third line of eq.
(\ref{secdir}) vanish for the "single-invariant" fields with
$\mathfrak{G}=0$ under consideration.

Thus, we find for the infrared limit of the polarization operator in
the magnetic- or electric-like field in the momentum representation,
$\Pi^{\rm IR}_{\mu\tau}(k,p)=\delta(k-p)\Pi^{\rm
IR}_{\mu\tau}(k),$\bee \label{fourier} \Pi^{\rm IR}_{\mu\tau}(k)=
\left(\frac{\rmd\mathfrak{L}(\mathfrak{F},0)}{\rmd
\mathfrak{F}}(\delta_{\mu\tau}k^2-k_\mu k_\tau )
+\frac{\rmd^2\mathfrak{L}(\mathfrak{F},0)}{\rmd\mathfrak{F}^2}(F_{\mu\alpha}k_\alpha)(F_{\tau\beta}k_\beta)\right.
+\nonumber\\\nonumber\\+\left.\left.\frac{\partial^2\mathfrak{L}(\mathfrak{F},
\mathfrak{G})}{\partial\mathfrak{G}^2}\right|_{\mathfrak{G}=0}
(\tilde{F}_{\mu\alpha}k_\alpha)(\tilde{F}_{\tau\beta}k_\beta)\right).\eend
Here the scalar $\mathfrak{F}$ and the tensors $F, \tilde{F}$ are
already set to be  space- and time-independent. By comparing this
with (\ref{pimunu}) we identify the eigenvalues of the polarization
operator in the infrared limit as
\bee\label{kappa}\left.\kappa_1(k^2,kF^2k,\mathfrak{F})\right|_{k\rightarrow
0}= k^2\frac{\rmd\mathfrak{L}(\mathfrak{F},0))}{\rmd\mathfrak{F}},
\nonumber\\\nonumber\\\left.\kappa_2(k^2,kF^2k,\mathfrak{F})\right|_{k\rightarrow
0} = k^2\frac{\rmd\mathfrak{L}(\mathfrak{F},0))}{\rmd\mathfrak{F}}-
(k\tilde{F}^2k)\left.\frac{\partial^2\mathfrak{L}(\mathfrak{F},\mathfrak{G})}
{\partial\mathfrak{G}^2}\right|_{\mathfrak{G}=0},\nonumber\\\nonumber\\
\left.\kappa_3(k^2,kF^2k,\mathfrak{F})\right|_{k\rightarrow 0}=
k^2\frac{\rmd\mathfrak{L}(\mathfrak{F},0))}{\rmd\mathfrak{F}}-
(kF^2k)\frac{\rmd^2\mathfrak{L}(\mathfrak{F},0)}
{\rmd\mathfrak{F}^2}.\eend This is the leading behavior  of the
polarization operator in the magnetic-like field near
zero-momentum point $k_\mu =0$. Every eigenvalue $\kappa_a$ turns
into zero quadratically when all the  momentum components
disappear. Thereby, eq. (\ref{gauge}) is proved.

For the sake of completeness, we give the same eqs. (\ref{kappa})
also in terms of the invariant variables 
 \bee\label{variables}H=
\sqrt{\mathfrak{F}+\sqrt{\mathfrak{F}^2+\mathfrak{G}^2}}\qquad
E=\sqrt{-\mathfrak{F}+\sqrt{\mathfrak{F}^2+\mathfrak{G}^2}}\eend
that are, respectively,  the  magnetic and electric fields in the
Lorentz frame, where these are parallel. Then, with the notation
${L}(H,E)=\mathfrak{L}(\mathfrak{F},\mathfrak{G})$ 
the coefficients in (\ref{kappa}) are :\bee\label{subst}
 \frac{\rmd\mathfrak{L}(\mathfrak{F},0)}{\rmd\mathfrak{F}}=
\frac{1}{H}\frac{\rmd L(H,0)}{\rmd H},\nonumber\\
 \frac{\rmd^2\mathfrak{L}(\mathfrak{F},0)}{\rmd\mathfrak{F}^2}=
 \frac 1{2\mathfrak{F}}\left(\frac{\rmd^2L(H,0)}{\rmd H^2}-\frac{~~\rmd L(H,0)}{H\rmd H}\right),\nonumber\\
 \left.\frac{\partial^2\mathfrak{L}(\mathfrak{F},
\mathfrak{G})}{\partial\mathfrak{G}^2}\right|_{\mathfrak{G}=0}=
\frac{1}{2\mathfrak{F}}\left. \left(\frac 1{E}\frac{\partial
L(H,E)}{\partial E}\right)\right|_{E=0}+
 \frac{1}{2\mathfrak{F}}\frac 1{H}\frac{~\rmd L(H,0)}{\rmd H}.\eend

 At this step we turn  to the special case of magnetic-like background
 and shall be sticking to it until the end of the present Subsection,
 keeping the extension of some results to the electric-like case
$\mathfrak{F}<0$  to the next Subsection C.

The dispersion curves $f_a(k_\perp^2)$ near the origin may be found
by solving equations (\ref{dispersion}) in the special frame with
the right-hand sides taken as (\ref{kappa}) and with eqs.
(\ref{special}) taken into account. This gives for the photons of
modes 2 and 3
\bee\label{linear2}f_2(k_\perp^2)=k_\perp^2\left(\frac{1-\mathfrak{L_F}}
{1-\mathfrak{L_{F}}+2\mathfrak{F}\mathfrak{L_{GG}}}\right), \eend
\bee\label{linear3}f_3(k_\perp^2)=k_\perp^2\left(1-\frac{2\mathfrak{F}\;
\mathfrak{L_{FF}}}{1-\mathfrak{L_{F}}}\right), \eend where we are
using the notations
$\mathfrak{L_{FF}}=\frac{\rmd^2\mathfrak{L}(\mathfrak{F},0)}
{\rmd\mathfrak{F}^2},\quad$
$\mathfrak{L_{F}}=\frac{\rmd\mathfrak{L}(\mathfrak{F},0))}{\rmd\mathfrak{F}},\quad
\mathfrak{L_{GG}}=\left.\frac{\partial^2\mathfrak{L}(\mathfrak{F},\mathfrak{G})}
{\partial\mathfrak{G}^2}\right|_{\mathfrak{G}=0}.$ As for mode 1,
the dispersion equation in the present approximation has only the
trivial solution $k^2=0$ that makes the vector potential
$\flat_\mu^{(1)}$ corresponding to it purely longitudinal, with no
electromagnetic field  carried by the mode. This is a nonpropagating
mode in the infrared limit (it is also nonpropagating within the
one-loop approximation beyond this limit; however,
massive-positronium  solutions in mode 1 do propagate \cite{ass}).

The unitarity condition (\ref{unitarity}), as applied to mode 2,
gives via the second equation in (\ref{kappa})
\bee\label{unitarity2}1-\mathfrak{L_F}+2\mathfrak{F}\mathfrak{L_{GG}}\geq
0. \eend 
Then,  from the  behavior of the dispersion curve (\ref{linear2})
and
the causality (\ref{causality}) 
it follows that \bee\label{unicaus}1-\mathfrak{L_F}\geq 0 \eend
and \bee\label{min2}\mathfrak{L_{GG}} \geq 0.\eend 
(Remind that for the magnetic-like  case under consideration one
has
$\mathfrak{F}>0$.) 

Analogously, the unitarity condition (\ref{unitarity}), as applied
to mode 3, gives via the third equation in (\ref{kappa}) again the
result (\ref{unicaus}). (This inequality also provides the
positiveness of the norm of the non-propagating mode 1.) Then from
the  behavior of the dispersion curve (\ref{linear3}) and the
causality (\ref{causality}) it follows that \bee\label{unitarity3}
1-\mathfrak{L_F}-2\mathfrak{F}\mathfrak{L_{FF}}\geq 0\eend and
\bee\label{LGG}\mathfrak{L_{FF}} \geq 0.\eend

Inequalities eq.(\ref{unicaus}),  eq.(\ref{unitarity3}) together
provide that all the three residues of the photon Green function in
the complex plane of $k_\perp^2$, the same as in the complex plane
of $(k_3^2-k_0^2)$, eq.(\ref{unitarity}), are also nonnegative
\bee\label{residue2}
1-\left.\frac{\partial\kappa_a(k^2,k_\perp^2,\mathfrak{F})}{\partial
k_\perp^2}\right|_{k_0^2-k_3^2=f_a(k_\perp^2)}\geq 0,\eend at least
in the infrared limit. We do not know whether this statement is
prescribed by general principles and therefore might be expected to
hold beyond this limit.

 Relations (\ref{min2}),  (\ref{LGG}) indicate that the extremum of the effective action at
$\mathfrak{G}=0$ (note that
$\left.(\partial\mathfrak{L})/\partial\mathfrak{G})\right|_{\mathfrak{G}=0}=0
$ due to P-invariance) is a minimum for any $\mathfrak{F}$ and that
the Lagrangian is a convex function of $\mathfrak{F}$ for any
$\mathfrak{F} > 0$ and of $\mathfrak{G}$ for $\mathfrak{G}=0$.

Relations (\ref{unitarity2}), (\ref{unicaus}), (\ref{unitarity3})
indicate positiveness of various dielectric and magnetic
permittivity constants that control electro- and magneto-statics
of charges and currents of certain configurations.
Eqs.(\ref{kappa}) imply that the quantities that are subject to
the inequalities (\ref{unitarity2}), (\ref{unitarity2}) and
(\ref{unitarity3}) are expressed in terms of  different infra-red
limits of the polarization operator eigenvalues as
\bee\label{from17a} 1-\mathfrak{L_F}=\lim_{k_\perp^2\rightarrow0}
\left(1-\frac{\left.\kappa_2\right|_{k_0=k_3=0}}{k_\perp^2}\right)\equiv\varepsilon_{\rm
tr}(0),\nonumber\\
1-\mathfrak{L_F}=\lim_{k_\perp^2\rightarrow0}
\left(1-\frac{\left.\kappa_1\right|_{k_0=k_3=0}}{k_\perp^2}\right)\equiv\left(\mu^{\rm
w}_{\rm tr}(0))\right)^{-1},\nonumber\\
1-\mathfrak{L_F}=\lim_{k_3^2\rightarrow 0}
\left(1-\frac{\left.\kappa_3\right|_{k_0=k_\perp=0}}{k_3^2}\right)\equiv\left(\mu^{\rm
pl}_{\rm long}(0)\right)^{-1}, \eend
 \bee\label{from17b}
1-\mathfrak{L_F}+2\mathfrak{F}\mathfrak{L_{GG}}=\lim_{k_3^2\rightarrow0}
\left(1-\frac{\left.\kappa_2\right|_{k_0=k_\perp=0}}{k_3^2}\right)\equiv\varepsilon_{\rm
long}(0),\eend
\bee\label{from17c}
1-\mathfrak{L_F}-2\mathfrak{F}\mathfrak{L_{FF}}=\lim_{k_\perp^2\rightarrow0}
\left(1-\frac{\left.\kappa_3\right|_{k_0=k_3=0}}{k_\perp^2}\right)\equiv\left(\mu^{\rm
pl} _{\rm tr}(0)\right)^{-1}.\eend It is demonstrated in {\bf
Appendix} that $\varepsilon_{\rm long}$ and $\varepsilon_{\rm tr}$
are dielectric constants responsible for polarizing the
homogeneous electric fields parallel and orthogonal to the
external magnetic field, which are produced, respectively, by
uniformly charged planes( sufficiently far from them), oriented
across the external magnetic field and parallel to it, see
eqs.(\ref{elfield}) and (\ref{elfieldtrans}). These are determined
by the eigenvalue $\kappa_2$, the virtual photons of the mode 2
being carriers of electrostatic force.

The quantity $\mu^{\rm w}_{\rm tr}(0)$ is the magnetic
permittivity  constant responsible for attenuation of the magnetic
field produced by a constant current concentrated on a line,
parallel to the external magnetic field, sufficiently far from the
current-carrying line, see eq.(\ref{remind}) with  $\mu(0)$
replaced by
 $\mu^{\rm w}_{\rm tr} (0)$ in it. The same quantity $\mu^{\rm w}_{\rm tr}(0)$
 governs the constant magnetic field of a plane current flowing along the external field.
 This magnetic permittivity  is determined by the mode 1.
 The other two magnetic permittivities, $\mu^{\rm
pl} _{\rm long}(0)$ and $\mu^{\rm pl} _{\rm tr}(0)$ are determined
by the mode 3. The permittivity $\mu^{\rm pl} _{\rm tr}(0)$ is
responsible for  remote attenuation of the  magnetic field
produced by a constant current, homogeneously concentrated on a
plane, parallel to the external magnetic field, and flowing in the
direction transverse to it, see eq.(\ref{planemag}). This magnetic
field is homogeneous and parallel to the external field. Finally,
permittivity $\mu^{\rm pl} _{\rm long}(0)$ is responsible for
remote attenuation of the  magnetic field produced by a constant
straight current, homogeneously concentrated on a plane,
transverse to the external magnetic field, see
eq.(\ref{planemag2}). This field is also homogeneous. Virtual
photons of the modes 1 and 3 are carriers of magneto-static force.

By using the  wordings "sufficiently far" and "remote" we mean
distances from the corresponding sources that essentially exceed a
characteristic length of an underlying microscopic theory, wherein
the linear response is formed. In a material medium that may be an
interatomic distance; in perturbative QED this is the electron
Compton length.

Relations (\ref{from17a}), (\ref{from17b}), (\ref{from17c}) mean
that the inequalities (\ref{unitarity2}), (\ref{unicaus}) and
(\ref{unitarity3}) signify the positiveness of all the
characteristic  permittivities of the magnetized vacuum, which was
derived above on general basis. Besides, thanks to (\ref{from17a}),
there exists the equality between one dielectric and two (inverse)
magnetic permittivities\bee\label{equality}\varepsilon_{\rm
tr}(0)=\left(\mu^{\rm w}_{\rm tr}(0)\right)^{-1}=\left(\mu^{\rm
pl}_{\rm long}(0)\right)^{-1}.\eend The first equality here is a
direct consequence of the invariance under the Lorentz boost along
the magnetic field in the special frame (see eq. (\ref{1to2}) in
{\bf Section IV}) and can be extended to the permittivity functions
as defined in {\bf Appendix} by (\ref{mu}) and the right equation
(\ref{difconsts}), $\varepsilon_{\rm tr}(k_\perp^2)=\left(\mu^{\rm
w}_{\rm tr}(k_\perp^2)\right)^{-1}$.
\subsection{Electric-like background field}
In this Subsection we shall see how the inequalities
(\ref{unitarity2})--(\ref{LGG})  derived in the previous Subsection
are extended to the negative domain of the invariant $\mathfrak{F}$.

  Bearing in mind eqs. (\ref{special2}) we may solve again  dispersion equations (\ref{dispersion2})
   using eqs. (\ref{kappa}) to get  the photon dispersion curves in the electric-like background field in the
infrared approximation. For mode 2 this results in
\bee\label{disp2}k_0^2-k_3^2=k_\perp^2\left(1+\frac{2\mathfrak{FL_{GG}}}{1-\mathfrak{L_F}}\right),\eend
while  for mode 3 in
\bee\label{disp3}k_0^2-k_3^2=k_\perp^2\left(\frac{1-\mathfrak{L_F}}{1-\mathfrak{L_F}-2\mathfrak{FL_{FF}}}\right)\eend
(compare this with (\ref{linear2}), (\ref{linear3})). The unitarity
relation (\ref{unitarity}) applied to mode 2 leads to the inequality
(\ref{unicaus}). The causality condition (\ref{causality}), when
applied to (\ref{disp2}) requires that\bee\label{ineq}
\left(1+\frac{2\mathfrak{FL_{GG}}}{1-\mathfrak{L_F}}\right)^2\leq
\left(1+\frac{2\mathfrak{FL_{GG}}}{1-\mathfrak{L_F}}\right).\eend
This implies that the right-hand side of the inequality (\ref{ineq})
be positive and  thus the both sides can be divided on it. Then the
inequality (\ref{ineq}) becomes the inequality
(\ref{unitarity2})\bee
\left(1+\frac{2\mathfrak{FL_{GG}}}{1-\mathfrak{L_F}}\right)<1.\eend
In view of (\ref{unicaus}) this means that $2\mathfrak{FL_{GG}}<0.$
Once $\mathfrak{F}$ is negative for the electric -like case under
consideration now, we come again to the convexity condition
(\ref{min2}), now in the domain of negative $\mathfrak{F}$. By
applying the same procedure to mode 3 we quite analogously reproduce
eqs. (\ref{unitarity3}) and (\ref{LGG}).
\subsection{Energy-momentum conditions}
Apart from the  relations derived above, there is an extra relation
that does not include derivatives of the Lagrangian and follows from
the positiveness of the energy density of the background field. The
standard expression for the energy-momentum
tensor\bee\label{noether} T_{\mu\nu}(x)= -\frac{\partial L_{\rm
tot}}{\partial (\partial A_\alpha/\partial x_\nu)}\frac{\partial
A_\alpha}{\partial x_\mu}+\delta_{\mu\nu}L_{\rm tot}^{\rm sqr},\eend
$g_{00}=-1$, leads in the constant magnetic-like background, after
symmetrization, to
\bee\label{tmunuconst}T_{\mu\nu}=-
F^2~_{\mu\nu} (1-\mathfrak{L_F})+\delta_{\mu\nu}L_{\rm tot}.\eend
The trace of this tensor is $4(\mathfrak{L}-\mathfrak{FL_F})$. The
energy density in the special frame is
\bee\label{density}T_{00}=-L_{\rm tot}=
\mathfrak{F}-\mathfrak{L}.\eend Therefore, the condition
\bee\label{density2}\mathfrak{L}<\mathfrak{F}\eend should hold. Once
$T_{i0}=0$ the spacial part of the energy-momentum vector is zero,
hence the latter  is directed along the time.

 Now we
proceed by describing general restrictions imposed by the physical
requirement that the energy density of elementary excitations of the
magnetic-like  background (magnetized vacuum) be nonnegative ("weak
energy condition" in terms of Ref.
\cite{hawking})\bee\label{endensity}t_{00}\geq 0\eend and that their
energy-momentum flux density be non-spacelike ("dominant energy
condition" of Ref.
\cite{hawking}))\bee\label{poynting}t_{0\nu}^2\leq 0.\eend To this
end we have to define the energy-momentum tensor $t_{\mu\nu}(x)$ of
small perturbations of the background field by first defining their
 Lagrangian.

 The total effective
Lagrangian $L_{\rm tot}=-\mathfrak{F}+\mathfrak{L}$
 expanded near the background constant magnetic field contributes
 into the total action -- in view
  of the definition (\ref{Pi}) -- the following
 correction, quadratic in the small perturbation $a_\mu(x)$ above the
 background:  \bee\label{qudrcor}S_{\rm tot}^{\rm
 sqr}=\frac 1{2}\int a_\mu(x)\{-\left(\delta_{\mu\nu}\partial^2_\alpha-\frac\partial{\partial x_\mu}
\frac\partial{\partial
y_\nu}\right)\delta(x-y)+\Pi_{\mu\nu}(x,y)\}a_\nu(y)\rmd^4x\rmd^4y.\eend
The field intensity of the perturbation will be denoted as
$f_{\mu\nu}=\partial_\mu a_\nu - \partial_\nu a_\mu$. Using the
diagonal form of the polarization operator (\ref{pimunu}) we get in
the momentum representation \bee\label{nonlocal}L_{\rm tot}^{\rm
sqr}(k)=\frac 1{4}f^2+\frac1{4}\left(-\frac{\kappa_1}{k^2}f^2
+\frac{\kappa_1-\kappa_2}{2k\widetilde{F}^2k}(f\widetilde{F})^2+
\frac{\kappa_1-\kappa_3}{2k{F}^2k}(f{F})^2\right).\eend Here the
notations are used:
$(fF)_{\mu\nu}=f_{\mu\alpha}F_{\alpha\nu}=(Ff)_{\nu\mu},~$$
(fF)=(fF)_{\mu\mu}=(Ff)$,
$~f^2_{~\mu\nu}=f_{\mu\alpha}f_{\alpha\nu},~$
$f^2=f^2_{~\mu\mu}=-(f_{\mu\nu})^2$, and we have exploited the
relations $f^2=-2a_\mu(k^2\delta_{\mu\nu}-k_\mu k_\nu)a_\nu$,
$(fF)=2(aFk)$. This Lagrangian is nonlocal, since it depends on
momenta in a complicated way, in other words, it depends highly
nonlinearly on the derivatives with respect to coordinates. It
becomes local if we restrict ourselves to the infrared limit by
substituting eqs.(\ref{kappa}) into it. Then the quadratic
Lagrangian acquires the very compact form\bee\label{compact}L_{\rm
tot}^{\rm sqr}=\frac 1{4}f^2(1-\mathfrak{L_F})+\frac {1}{8}\left(
\mathfrak{L_{GG}}(f\widetilde{F})^2+
\mathfrak{L_{FF}}(f{F})^2\right).\eend This Lagrangian, quadratic in
the field $f_{\mu\nu}(x)$, does not contain its derivatives,
$F_{\mu\nu},\widetilde{F}_{\mu\nu},\mathfrak{L_F},\mathfrak{L_{GG}}$
and $\mathfrak{L_{FF}}$ being constants depending upon the
background field alone. It governs small-amplitude low-frequency and
low-momentum perturbations of the magnetized vacuum, free of/ or
created by small sources. It might be obtained also directly by
calculating the second derivative (\ref{Pi}) of the Lagrangian
defined on constant fields.

Once the  background is translation-invariant, there is a conserved
energy-momentum tensor $t_{\mu\nu}(x)$ of the field $f_{\mu\nu}$
provided by the  Noether theorem by considering variations of this
field. Applying the  definition  (\ref{noether}) to the field of
small perturbation $a_\mu$  and to its Lagrangian (\ref{compact}) we
get \bee\label{noether} t_{\mu\nu}(x)= -\frac{\partial L_{\rm
tot}^{\rm sqr}}{\partial (\partial a_\alpha/\partial
x_\nu)}\frac{\partial a_\alpha}{\partial
x_\mu}+\delta_{\mu\nu}L_{\rm tot}^{\rm sqr}=\nonumber\\
=-\frac{\partial a_\alpha}{\partial
x_\mu}\left(f_{\alpha\nu}(1-\mathfrak
{L_F})+\frac1{2}(f\widetilde{F})\mathfrak{L_{GG}}\widetilde{F}_{\alpha\nu}
+\frac1{2}(f{F})\mathfrak{L_{FF}}{F}_{\alpha\nu}\right)
+\delta_{\mu\nu}L_{\rm tot}^{\rm sqr}.\eend The Maxwell equations
for small sourceless perturbations of the magnetized vacuum are
\bee\label{maxwell}\frac{\delta L_{\rm tot}^{\rm sqr}}{\delta
a_\alpha}=\frac{\partial}{\partial x_\nu}\frac{\partial L_{\rm
tot}^{\rm sqr}}{\partial (\partial a_\alpha/\partial
x_\nu)}=\frac{-\partial}{\partial
x_\nu}\left(f_{\alpha\nu}(1-\mathfrak
{L_F})+\frac1{2}(f\widetilde{F})\mathfrak{L_{GG}}\widetilde{F}_{\alpha\nu}
+\frac1{2}(f{F})\mathfrak{L_{FF}}{F}_{\alpha\nu}\right)=0.\eend We
are going to use the standard indeterminacy in the definition of the
energy-momentum tensor to let it depend only on the field strength
$f_{\mu\nu},$ and not on its potential. To this end we  add the
quantity (the designation $\doteq$ below means "equal up to full
derivative") \bee\label{quantity}\frac{\partial L_{\rm tot}^{\rm
sqr}}{\partial (\partial a_\alpha/\partial x_\nu)}\frac{\partial
a_\mu}{\partial x_\alpha}\doteq-a_\mu\frac\partial{\partial
x_\alpha}\frac{\partial L_{\rm tot}^{\rm sqr}}{\partial (\partial
a_\alpha/\partial x_\nu)}=\nonumber\\=a_\mu\frac\partial{\partial
x_\alpha} \{(f_{\alpha\nu}(1-\mathfrak
{L_F})+\frac1{2}(f\widetilde{F})\mathfrak{L_{GG}}\widetilde{F}_{\alpha\nu}
+ \frac1{2}(f{F})\mathfrak{L_{FF}}{F}_{\alpha\nu}\}\eend to
(\ref{noether}), that disappears due to the Maxwell equations
(\ref{maxwell}), taking into account the antisymmetricity of the
expression inside the braces. Hence the energy-momentum tensor may
be equivalently written as\bee\label{en-mom} t_{\mu\nu}(x)=
-f^2_{~\mu\nu}(1-\mathfrak
{L_F})-\frac1{2}(f\widetilde{F})\mathfrak{L_{GG}}(f\widetilde{F})_{\mu\nu}
-\frac1{2}(f{F})\mathfrak{L_{FF}}(f{F})_{\mu\nu}+\nonumber\\
+\frac{\delta_{\mu\nu}}4\left( f^2(1-\mathfrak{L_F})+
\frac1{2}\mathfrak{L_{GG}}(f\widetilde{F})^2+
\frac1{2}\mathfrak{L_{FF}}(f{F})^2\right).\qquad\qquad\eend This
tensor is traceless, $t_{\mu\mu}=0$. It obeys the continuity
equation with respect to the {\em second~}
index\bee\label{contin}\frac{\partial t_{\mu\nu}}{\partial
x_\nu}=0\eend owing to the Maxwell equations (\ref{maxwell}). Hence,
the 4-momentum vector obtained by integrating $t_{0\mu}$ over the
spatial volume $\rmd^3x$ conserves in time.

Let us take (\ref{en-mom}), first, on the monochromatic -- with
4-momentum $k_\mu$ -- real solution of the Maxwell equations
(\ref{maxwell}) that belongs to the eigen-mode 3:
$f_{\mu\nu}^{(3)}=k_\mu\flat^{(3)}_\nu-k_\nu\flat^{(3)}_\mu$. One
has
$(f^{(3)}{F})_{\mu\nu}=\flat^{(3)}_\mu\flat^{(3)}_\nu-k_\mu(F^2k)_\nu$,
$(f^{(3)}{F})=-2(kF^2k)$,
$(f^{(3)})^2_{\;\mu\nu}=-k^2\flat^{(3)}_\mu\flat^{(3)}_\nu+k_\mu
k_\nu(kF^2k),$ $(f^{(3)})^2=2k^2(kF^2k)$,
$(f^{(3)}\widetilde{F})=0$. With the substitution
$f_{\mu\nu}=f_{\mu\nu}^{(3)}$ the Maxwell equation (\ref{maxwell})
is satisfied, when \bee\label{sat}\flat^{(3)}_\alpha\{ k^2
(1-\mathfrak {L_F})+(kF^2k)\mathfrak{L_{FF}}\}=0,\eend i.e.,
naturally, on the dispersion curve (\ref{linear3}) for mode 3. 
It is seen that the Lagrangian (\ref{compact}) disappears on the
mass shell of mode 3, $L_{\rm tot}^{{\rm sqr} (3)}=0$. Then, the
reduction of the energy momentum tensor (\ref{en-mom}) onto this
mode, $t^{(3)}_{\mu\nu}(x),$ should be written with its
$\delta_{\mu\nu}$ part
 dropped: \bee\label{T3}t^{(3)}_{\mu\nu}(x)= (1-\mathfrak
{L_F})(k^2\flat^{(3)}_\mu\flat^{(3)}_\nu-k_\mu k_\nu(kF^2k))
+(kF^2k)\mathfrak{L_{FF}}(\flat^{(3)}_\mu\flat^{(3)}_\nu-k_\mu(F^2k)_\nu)
.\eend Although we referred to the magnetic-like background above in
 this Subsection, all the equations written in it up
to now remain, as a matter of fact,  valid also for the
electric-like case. In the rest of this Subsection we actually
specialize to the magnetized vacuum, although the conclusions may be
readily extended to cover  the electrified vacuum, as well.
  When $\mathfrak{F}>0,$ in the special frame (see eqs. (\ref{b}) in
Appendix), it holds $\flat^{(3)}_0=0$, $(F^2k)_{0,3}=0,$
$(F^2k)_{1,2}=-2\mathfrak{F}k_{1,2}.$
Then, after omitting the positive common factor
$-(kF^2k)=2\mathfrak{F}k_\perp^2$, we get for energy-momentum
density vector \bee\label{T0mu} t^{(3)}_{0\nu }(x)=
k_0\{(1-\mathfrak {L_F})k_\nu + \mathfrak{L_{FF}}(F^2k)_\nu\}
\eend  It is convenient to write it in components (counted as
0,1,2,3 downwards)
\bee\label{components}t_{0\nu}^{(3)}=k_0\left(\begin{tabular}{c}$k_0(1-\mathfrak{L_F})
$\\$k_1(1-\mathfrak{L_F}-2\mathfrak{FL_{FF})}$\\$k_2(1-\mathfrak{L_F}-2\mathfrak{FL_{FF})}$
\\$k_3(1-\mathfrak{L_F})$\end{tabular}\right)_\nu.\eend
The positive definiteness of the energy density (\ref{endensity})
results again in the requirement that the inequality (\ref{unicaus})
be satisfied. The causality in the form of the dominant energy
condition (\ref{poynting}) makes us expect that  vector
(\ref{components}) should
be non-spacelike. 
 Now, from (\ref{components}) with the use of the dispersion
law
(\ref{linear3}) 
this condition becomes \bee\label{momsqr}
t_{0,\mu}^{(3)2}=k_0^2\{(k_3^2-k_0^2)(1-\mathfrak{L_F})^2
+k_\perp^2(1-\mathfrak{L_F}-2\mathfrak{FL_{FF}})^2\}=\nonumber\\= -
2\mathfrak{FL_{FF}}k_0^2k_\perp^2(1-\mathfrak{L_F}-2\mathfrak{FL_{FF}})
\leq 0.\eend The same operations, performed over the energy-momentum
tensor (\ref{en-mom}) taken on mode 2 result in\bee\label{momsqr2}
t_{0,\mu}^{(2)2} = 
-2\mathfrak{FL_{GG}}k_0^2k_\perp^2(1-\mathfrak{L_F}+2\mathfrak{FL_{GG}})
\leq 0.\eend

The fulfillment of (\ref{momsqr}),  (\ref{momsqr2}) is guaranteed by
the inequalities (\ref{unitarity2}), (\ref{min2})--
(\ref{unitarity3}) established 
in Subsection B. However, the inverse statement would be wrong: the
inequalities (\ref{momsqr}), (\ref{momsqr2}), derived in the present
Subsection
do not yet lead to (\ref{unitarity2}), (\ref{min2})--
(\ref{unitarity3}). This may indicate that pair of conditions
(\ref{unitarity}) (unitarity as the positivity of the residue) and
(\ref{group}) (causality as the boundedness of the group velocity),
used to derive the limitations (\ref{unitarity2}) --
(\ref{unitarity3}) of Subsection B, are together more restrictive
than the two principles (\ref{endensity}) (energy positiveness) and
(\ref{poynting}) (causality as non-spacelikeness of the
energy-momentum density), although the latter provide the fact that
when solving the Cauchy problem initial data have no influence on
what occurs outside their light cone. (This is proved in
\cite{hawking} within General Relativity context. In this connection
it is interesting to mention the observation in Refs. \cite{novello}
that the background field may be represented by an equivalent
effective metric tensor at least as far as dispersion equations are
concerned. That metric tensor may be apparently used for
representing the Lagrangian (\ref{compact}) in geometric form.)

\section{Testing Euler-Heisenberg and Born-Infeld Lagrangians} In
 the one-loop approximation of QED
the quantities involved can be calculated either using the
Euler-Heisenberg effective Lagrangian
$\mathfrak{L}=\mathfrak{L^{(1)}}$ as long as the infrared limit is
concerned 
 or, alternatively, the one-loop polarization operator calculated
in \cite{batalin} for off-shell photons -- within and beyond this
limit. In the infrared limit the photon-momentum-independent
coefficients in (\ref{kappa}) within one loop are the following
functions
 of the dimensionless magnetic field $b=eB/m^2$, where $e$ and $m$ are the electron
charge and mass:\bee\label{L_F}\mathfrak{L^{(1)}_F}
=\frac\alpha{2\pi}\int_0^\infty\frac{\rmd t}{t}\exp\left({-\frac
t{b}}\right)\left(\frac{-\coth\;t}{t}+\frac1{\sinh^2t}+\frac2{3}\right),\eend
\bee\label{L_GG} 2\mathfrak{F}\mathfrak{L^{(1)}_{GG}}=\frac\alpha{3\pi}\int_0^\infty\frac{\rmd t}{t}\exp\left({-\frac
t{b}}\right)\left(\frac{-3\coth\;t}{2t}+\frac3{2\sinh^2t}+t\coth t\right),\eend
\bee\label{L_FF}  2\mathfrak{F}\mathfrak{L^{(1)}_{FF}}=
\frac\alpha{3\pi}\int_0^\infty\frac{\rmd t}{t}\exp\left({-\frac
t{b}}\right) \left(\frac{3\coth\;t}{2t}- \frac{t\coth t}{\sinh^2
t}+\frac3{2\sinh^2t}\right).\eend Here $\alpha=e^2/4\pi=1/137$ is
the fine-structure constant. Eq. (\ref{L_F}) turns to zero as
$\mathfrak{F}\sim b^2$, since the divergent linear in $\mathfrak{F}$
part of the one-loop diagram was absorbed in the course of
renormalization into $\mathfrak{L}_{\rm cl}$.
It can be verified
that the general relations (\ref{unitarity2})--(\ref{LGG})  ordained
by unitarity (\ref{unitarity}) and causality (\ref{causality}) to
the infrared limit  are obeyed by the one-loop approximation within
the vast range of the magnetic field values. However, due to the
known lack of asymptotic freedom in QED \cite{ritus2}, some of them
are violated for the exponentially strong fields. One can establish
the asymptotic behavior of (\ref{L_F}) - (\ref{L_FF}) in the limit
$b=eB/m^2\rightarrow\infty$\bee\label{asymp}
\mathfrak{L_F^{(1)}}\simeq \frac\alpha{3\pi}(\ln b-1.79),\qquad
2\mathfrak{F}\mathfrak{L_{GG}^{(1)}}\simeq
\frac\alpha{3\pi}(b-1.90),\qquad
2\mathfrak{F}\mathfrak{L_{FF}^{(1)}}\simeq \frac\alpha{3\pi}.\eend
 Thanks to the linearly growing \cite{skobelev} term in $\mathfrak{L_{GG}^{(1)}}$, for mode 2
 the positive-norm condition (the left relation in
(\ref{unitarity2})) is fulfilled for any $b$, and also the
dispersion curve (\ref{linear2}) goes outside the light cone, as
it is prescribed by the causality in the form of eq. (\ref{rest})
with $const =0$. However, the bracket in (\ref{linear2}) becomes
negative for $b>b_2^{\rm cr}=\exp\{1.79+3\pi/\alpha\}$, and mode 2
becomes a complex energy tachyon. For mode 3, the positive norm
condition (
 relation  (\ref{unicaus})) is fulfilled for  $b<b_2^{\rm cr}$. However, within the range
 exp$\{0.79+3\pi/\alpha\}=b_3^{\rm cr}<b<b_2^{\rm cr}$ the bracket in (\ref{linear3}) is negative,
 and mode 3 is a complex energy tachyon. For $b>b_2^{\rm cr}$ the dispersion curve (\ref{linear3})
 for mode-3 photon gets inside the light cone and becomes a super-luminal ghost with real energy
 and negative norm. An instability of the magnetized vacuum with respect to production of
 a constant field is associated with the imaginary energy at zero momentum. The elementary excitation
 with this property appears in mode 3 at a smaller threshold value, $b_3^{\rm cr}$,
 than in mode 2, $b_2^{\rm cr}$. The instability associated with mode-2 tachyons may
 lead to gaining the constant field with $\mathfrak{G}\neq 0$, since  the
 (pseudo)vector-potential $\flat^{(2)}_{\mu}$ (\ref{vectors}) carries
 an electric field component, parallel to the background magnetic
 field, whereas in $\flat^{(3)}_{\mu}$ this component is perpendicular to $\bf
 B$. It is interesting to note that, in spite of the instabilities and
 appearance of super-luminal excitations pointed above, the convexity properties
 (\ref{min2}), (\ref{LGG})
 are left intact under any magnetic field within one loop.

 The borders of stability of the magnetic field found here by
 analyzing the one-loop approximation are characterized by the
 large exponential $\exp\{1/\alpha\}$. It is much larger than the
 border found earlier \cite{2006} as the value where the
 mass defect of the bound electron-positron pair completely
 compensates the 2$m$ energy gap between the electron and
 positron, which is of the order of $\exp\{1/\sqrt\alpha\}.$
 These values are of the Planck scale.

 The situation is quite different for the Born-Infeld
 electrodynamics with its Lagrangian \bee\label{born} L_{\rm
 tot}=L^{\rm BI}=a^2\left(1-\sqrt{1+\frac{2\mathfrak{F}}{a^2}
 -\frac{\mathfrak{G}^2}{a^4}}\right)\eend
viewed upon as final, not
 subject to  further quantization. Here $a$ is an arbitrarily
 large  parameter with the dimensionality of mass squared. The
 correspondence principle (\ref{corresp}) is respected by eq. (\ref{born}).
 It does not contain field derivatives, hence all the infra-red limits encountered in this paper
 should be understood as exact values, for instance, going to the limit is unnecessary
 in (\ref{from17a}), (\ref{from17b}), (\ref{from17c}). The
 Lagrangian (\ref{born}) was derived long ago \cite{born} basing on very general
 geometrical principles of reparametrization-invariance, and
 besides it attracted much attention in recent decades thanks to the
 fact that it appears responsible for  the electromagnetic  sector of a
 string theory \cite{tseytlin} and thus is expected not to suffer from the lack of asymptotic freedom.
  For this reason our statement to follow that
 all the fundamental requirements established in Section 2 are
 obeyed in the Born-Infeld electrodynamics (\ref{born}) is
 instructive. We assume again that there is the constant and homogeneous
 magnetic-like external background and set
  $\mathfrak{G}=0$ after differentiation.
 Then, we get from
 (\ref{born})\bee\label{born2} 1-\mathfrak{L^{\rm
 BI}_F}=
 \left(1+\frac{2\mathfrak{F}}{a^2}\right)^{-\frac {1}{2}}\geq 0,\quad
 \mathfrak{L^{\rm BI}_{FF}}=a^{-2}
 \left(1+\frac{2\mathfrak{F}}{a^2}\right)^{-\frac3{2}}\geq 0,\quad \mathfrak{L^{\rm BI}_{GG}}=
 a^{-2}\left(1+\frac{2\mathfrak{F}}{a^2}\right)^{-\frac1{2}}\geq 0,\nonumber\\
 1-\mathfrak{L^{\rm BI}_F}+2\mathfrak{F}\mathfrak{L^{\rm BI}_{GG}}=
 \left(1+\frac{2\mathfrak{F}}{a^2}\right)^{\frac1{2}}\geq
0,\qquad 1-\mathfrak{L^{\rm BI}_F}-2\mathfrak{F}\mathfrak{L^{\rm
BI}_{FF}}=
 \left(1+\frac{2\mathfrak{F}}{a^2}\right)^{-\frac3{2}}\geq
0\qquad\qquad\eend where $\mathfrak{L^{\rm
 BI}}=L^{\rm
 BI}+2\mathfrak{F}.$ Thus, relations
 (\ref{unitarity2})--(\ref{LGG}) are all
 satisfied, hence there are neither ghosts, nor tachyons. The mode 1
 remains nonpropagating. As for modes 2 and 3, their dispersion
 curves coincide, since $f_2(k_\perp^2)=f_3(k_\perp^2)$ in (\ref{linear2}), (\ref{linear3}) due eqs.
 (\ref{born2}). This reflects the known absence of birefringence
 in the Born-Infeld electrodynamics \cite{plebanski}. Still,
 beyond the mass shell one has $\kappa_2\neq\kappa_3$, consequently the
 corresponding permeabilities  (\ref{from17a}), (\ref{from17b}), (\ref{from17c}) are different.
The same as in the one-loop QED, in the limit of large external
field there is a linearly growing contribution in $\kappa_2$, so
mode 2 dominates, the dielectric permeability (\ref{from17b})
behaving like the middle equation in
(\ref{asymp})\bee\label{behaving}\varepsilon_{\rm long}^{\rm
BI}(0) \simeq 2\mathfrak{F}\mathfrak{L_{GG}^{\rm BI}}\simeq \frac
B{a}\eend with the identification $a=(3\pi/\alpha)B_0$, where $B=
m^2/e=4.4\times 10^{14}$ Gauss is the characteristic field
strength in QED.

If we include the electric-like case we shall see that eqs.
(\ref{born2}) are all fulfilled within the interval
$-(2\mathfrak{F}/a) < \mathfrak{F} <\infty$, at the border of which
the Lagrangian (\ref{born}) becomes imaginary (recall that
$\mathfrak{G}=0$.)
\section{General basis for Adler's selection rule}
 There is an important statement that the dispersion in mode-2 photon is
 stronger than that in mode 3 throughout the range of continuity of
 the dispersion curves
 , i.e. $f_2(k_\perp^2) <  f_3(k_\perp^2)$ there.
 This statement holds within the one-loop approximation, where this range is $0<k_0^2-k_3^2<4m^2$, for all
 external fields and is crucial for establishing the
 kinematical selection rules for the photon splitting process \cite{adler}.
 In approximation-independent way this
 statement in the infrared limit
 might be expressed, following
eqs. (\ref{linear2}),  (\ref{linear3})
as\bee\label{which}\mathfrak{L_{GG}}-\mathfrak{L_{FF}}\geq
2\mathfrak{F}\frac{\mathfrak{L_{GG}}\mathfrak{L_{FF}}}{1-\mathfrak{L_{F}}}.\eend
Bearing in mind that the quantities $\mathfrak{L_{F}},
\mathfrak{L_{FF}}, \mathfrak{L_{GG}}$ are of the order of the fine
structure constant, this may be simplified just to
\bee\label{simlified}\mathfrak{L_{GG}} > \mathfrak{L_{FF}}.\eend We,
however, do not know whether this statement, simple as it is, can be
deduced from any fundamental principle. We can argue, nevertheless,
that the inequality $f_2(k_\perp^2) < f_3(k_\perp^2)$, once
fulfilled in the one-loop approximation (small $\alpha$), or at
least in the infrared limit, or at least for small magnetic field,
will remain valid for any $\alpha$, any momentum and any field. In
other words, dispersion curves of modes 2 and 3, considered as
functions of any of these parameters, cannot intersect, except in
the point $k_\mu=0$. Indeed, if they did, i.e. if the equality
$f_2(k_\perp^2) = f_3(k_\perp^2)$ might hold for a given choice of
$\alpha$, $B$, and $k_\mu\neq 0$, it would follow from
(\ref{dispersion}) and (\ref{law}) that also \bee\label{faul}
\kappa_2=\kappa_3 \qquad (? ?)\eend would be true on the mass shell
for the same choice. This degeneracy does take place at zero
momentum due to the  property (\ref{gauge}), also in the free case
$\alpha=0$, where all $\kappa_a$'s are zero, and in the
no-external-field case, where the isotropy of the vacuum is
expressed as $\kappa_1=\kappa_2=\kappa_3$, but would imply an extra
symmetry in the case of nonzero momentum.

Before discussing what sort of symmetry this might be, we dwell on
other   degeneracies  of the polarization operator - the ones that
are due to the residual Lorentz invariance left after the magnetic
field is imposed. These are the invariance under rotations about
the magnetic field direction (when ${\bf k}_\perp =0$) and under
Lorentz boosts along the magnetic field (when $k_3=k_0=0$). In the
limit ${\bf k}_\perp =0$ the  eigenvectors $\flat_\mu^{(1,3)}$
(\ref{vectors}), when normalized, turn into two unit 2-vectors
lying in the plane orthogonal to the magnetic field and orthogonal
to each other. They transform through each other under rotations
in this plane, while $\flat_\mu^{(2)}$ remains intact  (see
\cite{2008} or eq. (\ref{b}) in {\bf Appendix} for the explicit
form of the eigenvectors in the special frame to make sure of this
fact). Hence, referring to the representation (\ref{pimunu}), the
isotropy of the polarization tensor in this plane is expressed as
\bee\label{1to3}\left.\kappa_1\right|_{{\bf
k}_\perp=0}=\left.\kappa_3\right|_{{\bf k}_\perp=0}.\eend This
degeneracy provides that the virtual longitudinally directed
photons of modes 2 and 3, whose electric fields are lying in the
plane orthogonal to the magnetic field and are transverse to each
other, may be linearly combined to form two counter- and clockwise
transversely polarized eigenmodes. In the meanwhile the mode 2 is
a longitudinally polarized virtual eigenwave directed along the
magnetic field that corresponds to the quite different eigenvalue
$\kappa_2$.

In the other limiting case of $k_3=k_0=0$, quite analogously, the
eigenvectors $\flat_\mu^{(1,2)}$ (\ref{vectors})  turn after
normalization into two unit mutually orthogonal 2-vectors lying in
the hyperplane  $(k_3, k_0)$. They transform through each other
under Lorentz boosts along the magnetic field, while
$\flat_\mu^{(3)}$ is unchanged. Hence, the isotropy of the
polarization tensor (\ref{pimunu}) in this hyperplane is expressed
as\bee\label{1to2}\left.\kappa_1\right|_{
k_{0,3}=0}=\left.\kappa_2\right|_{ k_{0,3}=0}.\eend Eqs.
(\ref{1to2}) and (\ref{1to3}) are certainly obeyed within the
one-loop approximation.

We now come back to the wouldbe equality (\ref{faul}). Noting that
$\flat_\mu^{(2)}$ in (\ref{vectors}) is a pseudovector, whereas
$\flat_\mu^{(3)}$ is a vector, we see that (\ref{faul}), if true,
would imply the on-shell degeneracy with respect to parity. The
transformation that interchanges the vectors $\flat_\mu^{(2,3)}$
is the discrete duality transformation $B\to\rmi E$, $E\to -\rmi
B$, $F_{\mu\nu}\leftrightarrow \tilde{F}_{\mu\nu}$ -- not to be confused with continual duality. 
If we complete the
definition of the 
duality transformation by requiring
that on-shell the photon momenta do not change under it 
 we find
that eq. (\ref{faul}) would express the invariance of the
polarization operator in the form (\ref{pimunu}) under the 
duality transformation. No such invariance holds in QED already
because  there is no magnetic charge carrier in it. (The effective
Lagrangian on the class of  constant fields is still
dual-invariant, since the scalars $\mathfrak{F}$ and
$\mathfrak{G}^2$ on which it depends are. The Born-Infeld
Lagrangian above shared the same property, but it was completed by
the on-shell invariance (\ref{faul}) of the polarization operator
as well, expressed as the absence of birefringence, since the
asymmetry between  virtual magnetic and electric charges
(electrons) does not lie in the basis of Born-Infeld theory). Eq.
(\ref{faul}) is not fulfilled in any known approximation, except
for the trivial situations listed above. We conclude that an
intersection of dispersion curves of modes 2 and 3 should be ruled
out as a completely unbelievable event.
\section{Discussion}
 In  the present
paper, for establishing obligatory properties of the effective
Lagrangian we exploited two general principles --  unitarity and
causality -- taken in the special form of the requirements of
nonnegativity of the residue (\ref{unitarity}) and of boundedness of
the group velocity (\ref{group}).  We feel it necessary to confront
this way of action  with other approaches.

Usually, consequences of causality are discussed referring to
holomorphic properties of the polarization operator (dielectric
permittivity tensor) that follow from the retardation of the linear
response and are expressed -- after being supplemented by certain
statements concerning the high-frequency asymptotic conditions -- as
the Kramers-Kronig (once-subtracted) dispersion relations. Although
the general proof of an analog of the Kramers-Kronig relation in a
background field is lacking from the literature, for the magnetized
vacuum the holomorphity of the polarization operator eigenvalues
$\kappa_a$ in a cut complex plane of the variable $(k_0^2-k_3^2)$
was established within the one-loop approximation \cite{annphys},
\cite{shabtrudy}, the probability of electron-positron pair creation
by a photon making the cut discontinuity. Nevertheless, as we could
see in Section 3, this approximation contradicts some consequences
of the causality. Thus, the knowledge of the holomorphic properties
is not enough to exploit the causality requirement at full.

More specifically the causality  is approached by studying what is
called "causal propagation". Here the Hadamard's method  of
characteristic surface (the wave front), across which the first
derivative of the propagating solution may undergo a discontinuity
is used. The propagation is causal if the normal vector to the
characteristic surface is time- or light-like. (This criterion looks
very close to the group velocity criterion (\ref{group}) appealed to
by us.) Certain conditions obtained in this way that should be
obeyed by the "structural function $H$", the knowing of which is
equivalent to the effective Lagrangian, may be found among numerous
relations in a scrupulous study of Jerzy Pleba\'{n}ski. It seems,
however, that inequalities (9.176) derived in his Lectures
\cite{plebanski}, relating to the general case $\mathfrak{F}\neq0$,
$\mathfrak{G}\neq0,$ and the subsequent formulae, relating to the
 null-field subcase,
$\mathfrak{F}=\mathfrak{G}=0,$ need to be supplemented by
consequences of some requirements intended to substitute for
unitarity or positiveness of the energy, not exploited in
\cite{plebanski}, before/in-order-that a comparison with our
conclusions might become possible.

 On the other hand, when considering the causal propagation the
 implementation \cite{hawking} of
 Dominant Energy Condition (DEC) (\ref{poynting}) completed by  Weak Energy
 Condition (WEC) (\ref{endensity}) is also popular. The first one implies that
 the causality is reassured while solving the Cauchy boundary
 problem. One might expect that these two conditions are equivalent:
 first one to the group-velocity boundedness, and second one, at least partially,
 to the unitarity as the completeness of the set positive-energy
 states. The implementation of DEC and WEC to the problem of
 elementary excitations over the magnetized vacuum undertaken in Subsection C of Section
 II has indicated, however, as we already discussed it in that Subsection, that these
two conditions lead to somewhat weaker conclusions than the ones
that followed in Subsection B from
 imposing the conditions (\ref{unitarity}), (\ref{group}).

 We conclude by the remark that previously the appeal to the group
 velocity has shown its fruitfulness in establishing the phenomenon
 of canalization of the photon energy along the external magnetic
 field \cite{nuovocim}, \cite{annphys} and the capture of
 gamma-quanta by a strong nonhomogeneous magnetic field of a pulsar
 \cite{nature}, \cite{ass}.

\section*{Appendix} Here we are going to reveal direct physical
meanings to the constants involved in eqs. (\ref{unitarity2}),
(\ref{unitarity3}) in terms of various long-wave limits of the
dielectric and magnetic permeability of the vacuum in external
magnetic field. Before doing it we have to define these notions
within the technique of
 eigenvalues of the polarization operator \cite{batalin}, \cite{shabtrudy} used
throughout the present paper. We shall stress that the
proportionality relation  between the electric induction 
and electric field strength for electrostatic case common in
homogeneous isotropic medium cannot be naively extended to the
vacuum with magnetic field even though only one polarization mode
is dealt with. On the contrary, different small-momentum limits of
one and the same scalar dielectric function serve polarization of
external electric charges of  different configurations.

For the sake of direct comparison with the case of electrodynamics
of homogeneous isotropic medium we shall first consider this case
using the corresponding version of the technique of
 eigenvalues of the polarization operator \cite{vivian}.

In any homogeneous background the (second pair of)  Maxwell
equations can be written in the form
\bee\label{inhomogeneous}(k^2g_{\mu\rho}-k_\mu
k_\rho)A^\rho(k)-\Pi_{\mu\rho}(k)A^\rho(k)=-j_\mu(k),\quad
(j_\mu(k)k^\mu)=0, \eend where $j_\mu(k)$ is the conserved
(4-transversal) external current and $\Pi_{\mu\rho}(k)$ is the
4-transversal polarization operator,
$k_\mu\Pi_{\mu\rho}(k)=\Pi_{\mu\rho}(k)k_\rho=0$. Define the
electric induction  $\bf d$ as\bee\label{induction}
d_n=\varepsilon_{nj}e_j 
\quad n,j=1,2,3,\eend where $\bf e$ is the electric field
\bee\label{electric1} e_n=-\rmi (k_0A_n-k_nA_0)\eend and
\bee\label{eps}
\varepsilon_{nj}=\delta_{nj}+\frac{\Pi_{nj}}{k_0^2}. \eend With
the definition of the magnetic induction \bee\label{magnetic}
b_n=-\rmi\epsilon_{nji}k_jA_i\eend the first pair of the Maxwell
equations $\bf kb$=0 and $\epsilon_{nji}k_je_i-k_0b_n=0$ is
satisfied as a consequence of the definitions (\ref{electric1}),
(\ref{magnetic}), while (\ref{inhomogeneous}) becomes the second
pair of the (linearized near the external field) Maxwell equations
for $\bf d$ and $\bf b$ with external source
\bee\label{secpair}-\rmi{\bf kd}= j_0,\qquad
-\rmi(\epsilon_{nji}k_jb_i+k_0d_n)=j_n\eend with no polarization
charges and currents explicitly involved. These equations are
valid in the regions, where  the fields produced by the sources
$j$ are small as compared to the external field. This form of the
Maxwell equations, wherein the magnetic field strength and
induction are not distinguished may be found in \cite{LL}. We
reserve the letter $h$ for the magnetic field produced by the same
currents, but with all the magnetization effects disregarded. The
magnetic permeability will be defined with respect to those
fields.

We shall also need below eq.(\ref{induction}) in the form
\bee\label{eps2}d_n=e_n+\rmi \frac{\Pi_{n\mu}A^\mu}{k_0},\eend
which follows from (\ref{electric1}), (\ref{eps}) and the
4-transverseness of the polarization tensor. The Fourier transform
is defined as\bee\label{fourier} D_{\mu\nu}(x)=\frac
1{(2\pi)^4}\int \exp ({\rm i}kx) D_{\mu\nu}(k)~{\rm d}^4k,\quad
\mu,\nu=0,1,2,3.\eend The equation giving the 4-potential in terms
of the external 4-current to be used throughout this Appendix
is\bee\label{4-pot} A_\mu(x)=\int
D_{\mu\nu}(x-y)j^\nu(y)\rmd^4y,\quad \mu,\nu=0,1,2,3.\eend Here
$x$ and $y$ are 4-coordinates, and $D_{\mu\nu}(x-y)$ is the photon
Green function in a magnetic field in the coordinate
representation.
 \subsection{Isotropic medium}
The most general covariant 4-transversal  polarization tensor of a
isotropic homogeneous medium \cite{fradkin} formed with the  use
of the 4-velocity $u_\mu$ of the medium  is diagonal \cite{vivian}
in the basis $a_\mu$, $c^{(n)}_\mu, n=1,2$
\bee\label{Pidiag}\Pi_{\mu\nu}(k)=
\kappa_1(k^2,(uk)^2)\sum_{n=1,2}\frac{c^{(n)}_\mu
c^{(n)}_\nu}{(c^{(n)})^2}+\kappa_2(k^2,(uk)^2)\frac{a_\mu
a_\nu}{a^2}\eend with \bee\label{a} a_\mu=u_\mu k^2-k_\mu
(uk),\qquad
a^2=k^2(k^2-(uk)^2), \qquad (au)=0, \eend 
and $c^{(1)}_\mu$  defined as any 4-vector orthogonal to the
hyperplane, where the two vectors $k_\mu$ and $a_\mu$ lie, whereas
$c^{(2)}_\mu\equiv\varepsilon_{\mu\nu\rho\lambda}c^{(1)}_\nu
a_\rho k_\lambda$ is also orthogonal to this hyperplane and,
besides, orthogonal to $c^{(1)}_\mu$. Thus  the four vectors
$k_\mu,\, a_\mu$, and $c^{(1,2)}_\mu$ make an orthogonal basis in
the Minkowski space. They are four  eigenvectors of the
polarization
operator\bee\label{eigena}\Pi_{\mu}^{~\nu}(k)c^{(1,2)}_\nu=\kappa_1
c^{(1,2)}_\mu,\nonumber\\
\Pi_{\mu}^{~\nu}(k)a_\nu=\kappa_2a_\mu,\qquad \Pi_{\mu}^{~\nu}(k)
k_\nu=0.\eend Only three basis vectors appear in the decomposition
(\ref{Pidiag}) because one eigenvalue is zero, according to
(\ref{eigena}). The four basis vectors $k_\mu,\, a_\mu$, and
$c^{(1,2)}_\mu$ are 4-vector potentials of the electromagnetic
eigen-waves. In the rest-frame of the medium ($u_\mu=\delta_{\mu
 0}$) (and arbitrary normalization) these may be taken  in the form -- the components are
counted downwards as $\nu= 0,1,2,3$ --
\begin{eqnarray}\label{b1}
c_\nu^{(1)}=\left(\begin{tabular}{c}0\\$k_3$\\0\\$-k_1$\end{tabular}\right),\quad
c_\nu^{(2)}=\left(\begin{tabular}{c}$0$\\$k_1k_2$\\$-(k_3^2+k_1^2)$\\$k_2 k_3$\end{tabular}\right),\nonumber\\
a_\nu=k^2\left(\begin{tabular}{c}1\\$0$\\$0$\\0\end{tabular}\right)-k_0\left(\begin{tabular}
{c}$k_0$\\$k_1$\\$k_2$\\$k_3$\end{tabular}\right),\quad
k_\nu=\left(\begin{tabular}{c}$k_0$\\$k_1$\\$k_2$\\$k_3$\end{tabular}\right)_\nu.
\end{eqnarray} The orientations of the corresponding electric and
magnetic fields, calculated basing on these vector-potentials, are
described in detail in \cite{vivian}. In the Lorentz frame, where
the medium is at rest,  mode 1 is transversely-polarized
electromagnetic wave, while mode 2 is purely longitudinal electric
wave, its magnetic field being equal to zero. The degeneracy
corresponding to the fact that there is a common eigenvalue
$\kappa_1$ for the two eigenvectors $c^{(1,2)}_\mu$ reflects the
axial symmetry of the problem, which in the rest frame reduces to
the symmetry under rotations around the direction of the photon
3-momentum $\bf k$. If the kinematical condition $k^2=(uk)^2$ is
fulfilled, additional degeneracy \bee\label{degeneracy}
\kappa_1(k^2,k^2)=\kappa_2(k^2,k^2),\qquad {\rm if}\quad (uk)\neq
0 \eend appears that reflects a symmetry, which in the rest frame
is spherical symmetry due to the disappearance of the direction
specialized by the photon 3-momentum: in this frame the above
kinematic condition becomes just ${\bf k}^2=0$. The above-said
relates to real (on-shell) photons of the eigenmodes and to
virtual (off-shell) photons, as well. The latter are subject to
two, generally different, dispersion equations\bee\label{dispeq}
k^2=\kappa_{1,2}(k^2,(uk)^2).\eend To see this consider the
Schwinger-Dyson equation for the photon Green function
$D_{\mu\nu}(k)$ in momentum space:\bee\label{SchD}
(k^2g_{\mu\rho}-k_\mu k_\rho)D^\rho\;_\nu
(k)-\Pi_{\mu\rho}(k)D^\rho\;_\nu (k)=g_{\mu\nu}-k_\mu
k_\nu/k^2.\eend After the substitution of (\ref{Pidiag}) its
solution is readily found to be\bee\label{D}D_{\mu\nu}
(k)=\frac{1}{k^2-\kappa_1}\sum_{n=1,2}\frac{c^{(n)}_\mu
c^{(n)}_\nu}{(c^{(n)})^2}+\frac 1{k^2-\kappa_2}\frac{a_\mu
a_\nu}{a^2}+ k_\mu k_\nu D^{\rm L}(k).\eend Here $D^{\rm L}(k)$ is
an arbitrary function, not determined by the Schwinger-Dyson
equation.

\subsubsection {Electrostatics in isotropic medium.} Consider the electrostatic
problem with a source comprised of charges that are at rest in the
rest frame of the medium: \bee\label{statcharge} j_\nu
(k)=\delta_{\nu 0} \delta(k_0)q({\bf k}), \quad (jk)=0.\eend Then
the field produced by this static source is given by the
vector-potential
\begin{eqnarray}\label{potpoint} \hspace{-3cm} A_\mu({\bf x})=
\frac{1}{(2\pi)^3}\int D_{\mu 0}(0,{\bf k})\exp(-\rmi{\bf
kx})q({\bf k})\rmd^3k.
\end{eqnarray} Here zero stands for the $k_0$-argument of the
photon propagator.  Among the eigenvectors (\ref{b1}) there is
only one, whose zeroth component survives the substitution
$k_0=0$. It is $a_\nu$. For this reason only the second term in
(\ref{D}) contributes to (\ref{potpoint}). Spatial components of
$a_\nu$ are zero in this limit. Therefore,\bee\label{A00} A_0({\bf
x})=\frac{1}{(2\pi)^3}\int\frac{\rme^{-\rmi{\bf kx}}q({\bf k})
\rmd^3k}{{\bf k}^2-\kappa_2({\bf k}^2,0)},~~A_{1,2,3}({\bf
x})=0.\eend

Certainly, the static potential has only its zeroth component
different from zero and carries no magnetic field. Using this fact
in the definition of the induction (\ref{eps2}) we get for the
induction (\ref{induction}) corresponding to the potential
(\ref{A00}) (note that $a_n/k_0=-k_n$ and that $a^2=k^2{\bf k}^2)$
\bee\label{statind} {\bf d}=\rmi {\bf k}A_0(k)\left(1-\frac
{\kappa_2({\bf k}^2,0)}{{\bf k}^2}\right)={\bf e} \varepsilon({\bf
k}^2),\eend where\bee\label{scalepsilon}\varepsilon({\bf
k}^2)=1-\frac {\kappa_2({\bf k}^2,0)}{{\bf k}^2}\eend is the
static dielectric permittivity with spacial dispersion, equal to
the inverse refraction index squared, $\varepsilon({\bf
k}^2)=n_2^{-2},$ defined on the mass shell as (\ref{refrindex}).
The field strength and induction are parallel in the momentum
space, but
 in the configuration space,\bee\label{e3} {\bf
e(x)}=\frac{\rmi}{(2\pi)^3}\int {\bf k}\frac{\rme^{-\rmi{\bf
kx}}q({\bf k}) \rmd^3k}{{\bf k}^2\varepsilon (\bf k^2)},\eend
\bee\label{d3}{\bf d(x)}=\frac{\rmi}{(2\pi)^3}\int {\bf
k}\frac{\rme^{-\rmi{\bf kx}}q({\bf k}) \rmd^3k}{\bf k^2}\eend
they, generally, are not, except, for instance, spherical charge
distribution, $q({\bf k})=q({\bf k^2})$ or when considered far
from the charges -- where one may take $q({\bf k}\approx q({0}),$
-- and some other special cases.

Let us examine, next, a homogeneously charged, infinitely extended
plane, say the ({\em 1,2})-coordinate plane. This corresponds to
the choice $q({\bf k})=(2\pi)^2\rho\delta^2({\bf k_\perp})$ with
the 2-vector $\bf k_\perp$ lying in the chosen plane and $\rho$
being a constant surface charge density. Then \bee\label{esurf}
e_3(x_3)=\frac{\rmi\rho}{2\pi}\int\frac{k_3\rme^{-\rmi{k_3x_3}}\rmd
k_3}{{ k_3}^2\varepsilon (k_3^2)}, \quad e_{1,2}({\bf x})=0.\eend
If $|x_3|$ is large, only small $|k_3|$ contribute. Then we get
for the electric field the asymptotic expression
\bee\label{A0orth2} e_3({x_3})\approx\frac{\rmi\rho}{2\pi}\frac
1{\varepsilon(0)} \int\frac{\rme^{-\rmi{k_3x_3}}k_3\rmd k_3}{{
k_3}^2}\eend that is $1/\varepsilon$ multiplied by the field of
the charged plane without the  polarization taken into account.
The latter in the present case coincides with the induction
(\ref{d3}). To define the integral in the infra-red region one may
introduce a regularizing mass parameter $\mathfrak{m}>0$ and let
it tend to zero afterwards. (This gives the same result as the
causal shift of the pole $k_0^2-{\bf k}^2+\rmi 0$ in the photon
Green function). Then the field is
\bee\label{without}e_3({x_3})\approx\lim_{\mathfrak{m}\rightarrow
0} \frac{-\rho}{2\pi}\frac 1{\varepsilon(0)}\frac{\rmd}{\rmd x_3}
\int\frac{\rme^{-\rmi{k_3x_3}}\rmd k_3}{{
k_3}^2+\mathfrak{m}^2}=\lim_{\mathfrak{m} \rightarrow 0}\frac
{-\rho}{2\varepsilon(0)}\frac{\rmd}{\rmd x_3}\frac{\exp
(-|x_3|\mathfrak{m})}{\mathfrak{m}}. \eend Finally, in the
isotropic medium the electric field of a charged plane parallel to
the co-ordinate plane ({\em 1,2}) at large distance from this
plane is a constant field pointing to the plane:
\bee\label{elfield}
e_3(x_3)\approx\frac{\rho}{2\varepsilon(0)}{\rm sgn}(x_3),\eend
where sgn$(x)=\pm 1$ for $x\gtrless 0$. We have reproduced this
well-known result to stress that it is direction-independent: it
gives the electric field, orthogonal to the chosen charged plane,
as a function of the distance from that plane by a universal
formula, independent of the orientation of the plane. We shall see
in the next subsection, how this result will be modified in the
magnetized vacuum.

\subsubsection{Magneto-statics of isotropic medium.} Now consider the magneto-static
problem with the source corresponding to a constant current
flowing in the special frame along the direction {\em
3.}\bee\label{current} j_\mu =\delta_{\mu 3}j({\bf
k_{\perp}})\delta(k_0)\delta(k_3), \qquad (kj)=0,\eend where $\bf
k_{\perp}$ is the two-component momentum in the ({\em 1,2})-plane.
It produces the 4-vector potential
\begin{eqnarray}\label{potpoint2} \hspace{-3cm} A_\mu({\bf
x}_{\perp})= \frac{1}{(2\pi)^3}\int D_{\mu 3}(0,0,{\bf
k}_{\perp})\exp(-\rmi{\bf k_{\perp} x_\perp})j({\bf
k_\perp})\rmd^2k_\perp,
\end{eqnarray} where the zeros stand for the $k_0$- and $k_3$-arguments of the
photon propagator. Among the eigenvectors (\ref{b1}) there is only
one, whose third component survives the substitution $k_0=0$. It
is $c^{(1)}_\nu$. For this reason only the first term in (\ref{D})
with $n=1$ contributes to (\ref{potpoint2}). The $(\nu\neq 3)$-
components of $c^{(1)}_\nu$ are zero in this limit.
Therefore,\bee\label{A3} A_3({\bf
x_\perp})=\frac{1}{(2\pi)^3}\int\frac{\rme^{-\rmi{\bf k_\perp
x_\perp}}j({\bf k_\perp}) \rmd^2k_\perp}{{ k_\perp}^2-\kappa_1(
k_\perp^2,0)},~~A_{0,1,2}({ x_\perp})=0.\eend

This  4-potential has only its third component different from zero
and carries no electric field. The magnetic induction, formed with
the use of this 4-potential according to eq. (\ref{magnetic})
\bee\label{wire1} b_n({\bf x_\perp})=\frac {-\rmi\epsilon_{nm\rm
\it 3}}{(2\pi)^3}\int \frac{\rme^{-\rmi \bf k_\perp x_\perp}j({\bf
k_\perp})k_m\rmd^2 k_\perp}{
k_\perp^2-\kappa_1(k_\perp^2,0)
}\eend differs from the magnetic field $h^{\rm non}_n({\bf
x_\perp})$ produced by the same current (\ref{current}) in the
absence of the medium (i.e. when $\kappa_1=0$) \bee\label{wire0}
h^{\rm non}_n({\bf x_\perp})=\frac {-\rmi\varepsilon_{nm\rm \it
3}}{(2\pi)^3}\int \frac{\rme^{-\rmi \bf k_\perp x_\perp}j({\bf
k_\perp})k_m\rmd^2 k_\perp}{k_\perp^2
}\eend by the factor in the integrand \bee\label{mu} \mu({
k^2_\perp})=\left(1-\frac{\kappa_1( k^2_\perp ,0)}{
k^2_\perp}\right)^{-1}\eend to be identified as magnetic
permeability. Its long-wave limit
$\mu(0)=\left(\left.1-(\kappa_1({\bf k^2} ,0)/{\bf
k^2})\right|_{{\bf k}^2=0}\right)^{-1}$ serves  the asymptotic
behavior of magnetic field $\bf b(x_\perp),$ $\bf |x_\perp
|\rightarrow\infty$ produced by the current, which flows along the
axis $x_3$ and whose density decreases in the orthogonal plane
({\em 1,2}) away from the origin sufficiently fast,
$j(0)\neq\infty$ in (\ref{current})
- otherwise the integral might  be infra-red-divergent. 
 In this case \bee\label{w} b_n({\bf x_\perp})=\frac {-\rmi\epsilon_{nm\rm
\it 3}}{(2\pi)^3}\int \frac{\rme^{-\rmi \bf k_\perp x_\perp}\mu
(k_\perp^2)j({\bf k_\perp})k_m\rmd^2 k_\perp}{k_\perp^2}\approx
\frac {-\rmi\epsilon_{nm\rm \it 3}j(0)\mu (0)}{(2\pi)^3}\int
\frac{\rme^{-\rmi \bf k_\perp x_\perp}k_m\rmd^2 k_\perp}{
k_\perp^2}\eend Eq. (\ref{w}) also covers the case of the current,
 flowing along an infinitely thin cylindric rectilinear wire, with $j({\bf
 k_\perp})$ being taken as  $j({\bf
 k_\perp})=2\pi J$, where $J$ is the total constant current. Then
\bee\label{remind}b_n({\bf x_\perp})\approx\frac {-\rmi J\mu (0)
\epsilon_{nm\rm \it 3}}{(2\pi)^2}\int \frac{\rme^{-\rmi \bf
k_\perp x_\perp}k_m\rmd^2 k_\perp}{k_\perp^2}=\nonumber\\=\frac
{-\rmi J\mu (0)\epsilon_{nm\rm \it 3}}{(2\pi)^2{|\bf
x_\perp}|}\int \frac{\rme^{-\rmi \bf k_\perp
\hat{x}_\perp}k_m\rmd^2 k_\perp}{k_\perp^2}=\frac {-J\mu
(0)}{2\pi}\frac{\epsilon_{nm\rm \it 3}{\bf\hat{x}}_m}{|{\bf
x_\perp}|},\eend where $\hat{\bf x}_\perp$ is the unit vector
along the transverse coordinate $\hat{\bf x}_\perp=\frac{\bf
x_\perp}{\bf |x_\perp|}.$ To make sure that the last integration
is correct note that the projection of the integral onto the
direction orthogonal to $\bf x_\perp$ \bee\label{proj}
\int_0^\infty \rmd k \int_0^{2\pi} \rmd \phi\sin\phi\rme^{ -\rmi
k\cos\phi}=0\eend disappears due to the angle integration. It
could not be otherwise, since $\bf x_\perp$ is the only vector in
the integrand, hence the integral, which is a vector, and not a
pseudovector, cannot help being parallel to it. The projection on
$\bf x_\perp$ is \bee\label{proj2}\int_0^{2\pi} \rmd
\phi\int_0^\infty \rmd k \cos\phi\rme^{ -\rmi k\cos\phi}=-\rmi
2\pi\eend Certainly, the well-known relation (\ref{remind}) does
not, in the present case of isotropic medium, depend on the
specific choice of the direction of the current along the axis
$x_3$ made above.

\subsection{Magnetized vacuum} In the present subsection we refer
to the special frame and find it more convenient to list the
arguments in the eigenvalues $\kappa_a$ in (\ref{photon2}) in a
different order, also without indicating the dependence on the
magnetic field:
\bee\label{order}\kappa_a(k^2,-kF^2k,\mathfrak{F})=\bar{\kappa}_a(k_0^2,k_3^2,k_\perp^2)\eend
The first three (meaningful) 4-eigenvectors (\ref{vectors}) of the
polarization tensor $\Pi_{\mu\nu}$  take in the special frame (up to
the normalization, which we chose differently here) the form - the
components are counted downwards as $\nu= 0,1,2,3$ -
\begin{eqnarray}\label{b}
\flat_\nu^{(1)}=k^2\left(\begin{tabular}{c}0\\$k_1$\\$k_2$\\0\end{tabular}\right)_\nu-
k_\perp^2\left(\begin{tabular}{c}$k_0$\\$k_1$\\$k_2$\\$k_3$\end{tabular}\right)_\nu,\quad
\flat_\nu^{(2)}=\left(\begin{tabular}{c}$k_3$\\0\\0\\$k_0$\end{tabular}\right)_\nu,\quad 
\flat_\nu^{(3)}=\left(\begin{tabular}{c}0\\$k_2$\\$-k_1$\\0\end{tabular}\right)_\nu.\eend
Their lengths are
\bee\label{lengths}(\flat^{(1)})^2=-k^2k_\perp^2(k_0^2-k_3^2),\quad
(\flat^{(2)})^2=-(k_0^2-k_3^2), \quad (\flat^{(3)})^2=-k_\perp^2.
\end{eqnarray}
\subsubsection {Electrostatics of magnetized vacuum.} Consider the electrostatic problem
with source (\ref{statcharge}) comprised of charges that are at
rest in the special frame.  Then (\ref{4-pot}) results again in
(\ref{potpoint}).

Among the eigenvectors (\ref{b}) there is only one, whose zeroth component 
survives the substitution $k_0=0$. It is $\flat_\nu^{(2)}$. This
implies that out of the four ingredients of the general
decomposition of the photon propagator (\ref{photon2}) only the
term with $a=2$,
$D_2(k)\flat_{\mu}^{(2)}\flat_{\nu}^{(2)}/(\flat^{(2)})^2$,
participates in (\ref{potpoint}), i.e. only  mode-2 (virtual)
photon may be a carrier of electro-static interaction, and not
photons of modes 1,2, nor the purely gauge mode 4. Bearing in mind
that $(\flat^{(2)})^2=k_3^2-k_0^2$, we have\bee\label{A0} A_0({\bf
x})=\frac{1}{(2\pi)^3}\int\frac{\rme^{-\rmi{\bf kx}}q({\bf
k})\rmd^3k}{{\bf
k}^2-\bar{\kappa}_2(0,k_3^2,k_\perp^2)},~~A_{1,2,3}({\bf
x})=0.\eend Here $k_\perp^2=k_1^2+k_2^2$. Thus, the static charge
gives rise
to electric field only, as it might be expected. 
Equation for electric field (\ref{electric1}) corresponding to the
potential (\ref{A0}) can be represented as\bee\label{electric}{\bf
e}({\bf x})=\frac{1}{(2\pi)^3}\int\frac{\rmi{\bf k}
\rme^{-\rmi{\bf kx}}q({\bf k})\rmd^3k}{{\bf
k}^2-\bar{\kappa}_2(0,k_3^2,k_\perp^2)}=\nonumber\\
=\frac{1}{(2\pi)^3}\int\frac{\rmi{\bf k}\rme^{-\rmi{\bf kx}}q({\bf
k})\rmd^3k}{\varepsilon (k_3^2,k_\perp^2){\bf k}^2},\eend where
\bee\label{prop}\varepsilon
(k_3^2,k_\perp^2)=1-\frac{\bar{\kappa}_2(0,k_3^2,k_\perp^2)}{{\bf
k}^2}\eend is the coefficient -- to be understood as the
dielectric function -- of  proportionality between the (Fourier
transforms of) electric field in the magnetized vacuum and that
with the vacuum polarization disregarded
\bee\label{elinduction} {\bf e}_{\rm non}({\bf
x})=\frac{1}{(2\pi)^3}\int\frac{\rmi{\bf k}\rme^{-\rmi{\bf
kx}}q({\bf k})\rmd^3k}{{\bf k}^2}.\eend It is equal to the inverse
of the refraction index squared (\ref{refrindex}). Eq.
(\ref{elinduction}) does not coincide with the induction, ${\bf
e}_{\rm non}({\bf x})\neq {\bf d}({\bf x}),$ defined as
(\ref{induction}) or (\ref{eps2}), which is not, generally,
parallel with the electric field already in the momentum
space\bee\label{statind2} d_n=\rmi k_nA_0({\bf
k})\left(1-\delta_{n3}\frac {\bar{\kappa}_2(0,k_3^2, k_\perp^2)}{
k_3^2}\right)=\rmi \frac{k_nq({\bf
k})}{(2\pi)^3}\frac{\left(1-\delta_{n3}\frac{\bar{\kappa}_2(0,k_3^2,
k^2_\perp)}{ k_3^2}\right)}{{\bf
k}^2-\bar{\kappa}_2(0,k_3^2,k^2_\perp)}.\eend
 Once, unlike the
isotropic case (\ref{scalepsilon}), $\bar{\kappa}_2$ depends
separately on the two momentum squared components
$k_3^2,k_\perp^2$, there is
no universal, direction-independent 
static dielectric permeability. 
On the contrary, depending on the character of the external charge
distribution, one may speak of different dielectric functions,
which are, for instance, the two different long-wave limits of
(\ref{prop}) \bee\label{difconsts} \varepsilon_{\rm
long}(k_3^2)=1-\frac {\bar{\kappa}_2(0,k_3^2,0)}{k_3^2}, \qquad
\varepsilon_{\rm tr}(k_\perp^2)=1-\frac
{\bar{\kappa}_2(0,0,k_\perp^2)}{k_\perp^2}.\eend These two
dielectric functions control the potential far from the region
where the charges are located, in the directions across,
$\varepsilon_{\rm long}(k_3^2)$, and along the magnetic field,
$\varepsilon_{\rm tr}(k_\perp^2)$. One of situations of that sort,
namely the field of a point-like charge that  decreases with
different speeds along different directions following an
anisotropic Coulomb law, was studied in detail in \cite{prl2007},
\cite{2008} (see also \cite{sadooghi}) in the limit of large
magnetic field. If, on the contrary, the charge is not localized,
but distributed homogeneously along or across the magnetic field
the role of these two dielectric functions may be described more
definitely.

The first one, $\varepsilon_{\rm long}(k_3^2),$ is responsible for
polarization caused by the charge distribution, homogenous in the
direction orthogonal to the magnetic field, $q({\bf
k})=\delta^2({\bf k_\perp})\widetilde{q}(k_3)$. In particular, the
electric field strength of a plane ({\em 1,2}) charged  with a
constant surface density $\rho$,
$\widetilde{q}(k_3)=(2\pi)^2\rho,$ - oriented transversally to the
external magnetic field - is obtained from (\ref{electric}) as
\bee\label{elA0orth}
e_3(x_3)=\frac{\rmi\rho}{2\pi}\int\frac{k_3\rme^{-\rmi{k_3x_3}}\rmd
k_3}{{ k_3}^2-\bar{\kappa}_2(0,k_3^2,0)},\quad  {\bf
e_\perp(x)}=0.\eend If $|x_3|$ is large, only small $|k_3|$
contribute. Then, keeping the lowest term in the power series
expansion of $\bar{\kappa}_2(0,k_3^2,0)$ with respect to $k_3^2$,
we get -- in the same way as in the previous subsection -- that in
the vacuum, magnetized along the axis $x_3$, the electric field of
a charged plane parallel to the co-ordinate plane ({\em 1,2}) at
large distance from this plane is a constant field pointing to the
plane: \bee\label{elfield}
e_3(x_3)\approx\frac{\rho}{2\varepsilon_{\rm long}(0)}{\rm
sgn}(x_3).\eend In this case the induction is the same as the
electric field without the vacuum polarization, $\varepsilon_{\rm
long}(0){\bf e}({\bf x})={\bf e}_{\rm non}({\bf x})= {\bf d}({\bf
x})$.

The second dielectric function (\ref{difconsts}),
$\varepsilon_{\rm tr}(k_\perp^2)$, is responsible for polarization
caused by the charge distribution, homogenous in the direction
parallel to the magnetic field, $q({\bf
k})=\delta(k_3)\widetilde{\widetilde{q}}({\bf k_\perp})$. In
particular, the electric field of the coordinate plane ({\em 1,3})
charged with the constant surface density $\rho$,
$\widetilde{\widetilde{q}}({\bf k_\perp})=(2\pi)^2\rho
\delta({k_1})$ is obtained from (\ref{electric}) as
\bee\label{e2tr} e_2(x_2)=\frac{\rho}{2\pi}\int\frac{\rmi{
k_2}\rme^{-\rmi{ k_2x_2}}\rmd k_2}{{
k_2}^2-\bar{\kappa}_2(0,0,k_2^2)},\quad e_{1,3}({\bf x})=0. \eend
Keeping this time the lowest term in the power series expansion of
$\varepsilon(0,k_2^2)$ with respect to
$k_2^2$, we get far from the surface 
\bee\label{elfieldtrans}
e_2(x_2)\approx\frac{\rho}{2\varepsilon_{\rm tr}(0)}{\rm
sgn}(x_2).\eend Certainly, due to the axial symmetry of the
problem this result is basically the same for any charged plane
containing the vacuum magnetization direction {\em 3}, but differs
from (\ref{elfield}) in that it contains the different dielectric
constant. In the present case the induction is not
$\varepsilon_{\rm tr}(0){\bf e}({\bf x})={\bf e}_{\rm non}({\bf
x}),$ but, on the contrary,  coincides with the electric field
${\bf d}({\bf x})={\bf e}({\bf x})$. We, nevertheless, define the
dielectric permeability with respect to ${\bf e}_{\rm non}({\bf
x})$, and not with respect to ${\bf d}({\bf x}),$ the latter being
only introduced to give the Schwinger-Dyson equation the form of
the Maxwell equations in a medium.

 By confronting
eqs.(\ref{difconsts}) with (\ref{kappa}) we establish the
connection between the dielectric constants $\varepsilon_{\rm
long}(0),\;\varepsilon_{\rm tr}(0)$ with the quantities
(\ref{unitarity2}), and thus the nonnegativity of the
former\bee\label{connection} \varepsilon_{\rm
tr}(0)=1-\mathfrak{L}_\mathfrak{F}\geq 0, \quad \varepsilon_{\rm
long}(0)=1-\mathfrak{L}_\mathfrak{F}+2\mathfrak{F}\mathfrak{L}_{\mathfrak{GG}}\geq
0 .\eend

\subsubsection{Magneto-statics of magnetized vacuum.} Now consider the magneto-static
problem in the magnetized vacuum with the source (\ref{current})
corresponding to a constant current flowing in the special frame
along the direction of the external magnetic field {\em 3}. It
produces the same 4-vector potential as (\ref{potpoint2}), but
with the photon propagator given as (\ref{photon2}). Among the
three meaningful eigenvectors (\ref{b}) with $a=1,2,3$ there is
only one, whose third component survives -- after normalization --
the substitution $k_0=0$. It is $\flat^{(1)}_\nu$. Indeed, $(\flat
^{(1)})^2 = k^2k_\perp^2(k_3^2-k_0^2)={\bf k^2}k_\perp^2k_3^2$,
hence $\flat^{(1)}_3/\sqrt{(\flat ^{(1)})^2}=1$ after $k_3=0$ is
substituted. For this reason only the first term in
(\ref{photon2}) with $a=1$ contributes to (\ref{potpoint2}). The
$(\nu\neq 3)$-components of $\flat^{(1)}_\nu$ are zero in this
limit. Therefore, for the vector potential we have the equation
\bee\label{A31} A_3({\bf
x_\perp})=\frac{1}{(2\pi)^3}\int\frac{\rme^{-\rmi{\bf k_\perp
x_\perp}}j({\bf k_\perp}) \rmd^2k_\perp}{{
k_\perp}^2-\bar{\kappa}_1(0,0,k_\perp^2)},~~A_{0,1,2}({\bf
x_\perp})=0,\eend very similar to (\ref{A3}), but with the
external-magnetic-field-dependent eigenvalue $\kappa_1$, from
(\ref{eigen}), (\ref{photon2}). The magnetic induction, formed
with the use of this 4-potential according to eq. (\ref{magnetic})
differs from the magnetic field produced classically by the same
current (\ref{current}) in the non-magnetized vacuum (i.e. when
$\bar{\kappa}_1=0$) by the factor in the momentum space
\bee\label{mu} \mu^{\rm w}_{\rm tr}({
k^2_\perp})=\left(1-\frac{\bar{\kappa}_1(0,0, k^2_\perp)}{
k^2_\perp}\right)^{-1}\eend to be identified as magnetic
permeability. Its long-wave limit $\mu^{\rm w}_{\rm
tr}(0)=\left(\left.1-({\bar{\kappa}_1(0,0, k^2_\perp)/
k_\perp^2})\right|_{k_\perp^2=0}\right)^{-1}$ serves the
asymptotic behavior of magnetic induction $\bf  b(x_\perp),$ $\bf
|x_\perp |\rightarrow\infty$ produced by the current, which flows
along the external magnetic field and whose density decreases in
the orthogonal plane ({\em 1,2}) away from the origin sufficiently
fast, $j(0)\neq\infty$ in (\ref{current}). (This case includes the
straight-linear current of an infinitely thin wire.)  Now,
eqs.(\ref{w}) and (\ref{remind}) for the magnetic induction of the
current oriented along the axis {\em 3}, remain valid, but with
$\mu(k^2_\perp)$ and $\mu(0)$ replaced, respectively, by
 $\mu^{\rm w}_{\rm tr}(k^2_\perp)$ and $\mu^{\rm w}_{\rm tr} (0)$ in them.

The same quantity (\ref{w})  controls the magnetic induction of a
current also flowing parallel to the axis {\em 3}, but
homogeneously concentrated  on the plane that contains the
external magnetic field, say the ({\em 1,3})-plane, $j({\bf
k_\perp})=(2\pi)^2j\delta(k_1),$ where $j$ is a finite density of
current per unit length along the axis {\em 1}. Now, the potential
(\ref{A31}) becomes \bee\label{A3pl} A_3(
x_2)=\frac{j}{(2\pi)}\int\frac{\rme^{-\rmi k_2 x_2} \rmd k_2}{
k_2^2\mu^{\rm w}_{\rm tr}(k_2^2)},\eend and the corresponding
magnetic induction far from the surface, behaves (the infra-red
issue to be treated in the same way as in the electrostatic
problem of a charged plane (\ref{A0orth2}) above)
as\bee\label{planemaglong}b_1(x_2)\approx \frac {j\mu^{\rm w}_{\rm
tr}(0)}{2}{\rm sgn} (x_2).\eend

 The formulae hitherto obtained in this Subsubsection,
 however, are not  applicable to other directions of the
 current.

Let us, then, examine a constant current flowing in the magnetized
vacuum across its magnetic field, say along the axis {\em
1}\bee\label{transcur}j_\nu(k)=\delta_{\nu
1}\delta(k_0)\delta(k_1)j(k_3,k_2),\qquad (jk)=0.\eend The only
vector among (\ref{b}) that has nonzero component {\em 1}, when
$k_1=k_0=0$, and thus exclusively contributes into the photon
Green function (\ref{photon2}) is $\flat_\nu^{(3)}$. All the other
components of $\flat_\nu^{(3)}$ disappear in this limit.
Therefore, the potential produced by the current (\ref{transcur})
is \bee\label{A1} A_1(
x_2,x_3)=\frac{1}{(2\pi)^3}\int\frac{\rme^{-\rmi(k_2 x_2+k_3
x_3)}j(k_2,k_3) \rmd k_2\rmd
k_3}{k_2^2+k_3^2-\bar{\kappa}_3(0,k_3^2,
k_2^2)},~~A_{0,2,3}(x_2,x_3)=0.\eend An essential difference of
this expression from (\ref{A31}) or  (\ref{A3}) is that the axis
{\em 1} is not a symmetry axis. This is reflected in the fact that
$\bar{\kappa}_3(0,k_3^2, k_2^2)$ in (\ref{A1}) does not depend on
the combination $k_3^2+k_2^2,$ but contains the variables $k_2^2$
and $k_3^2$ separately. For this reason we have to further specify
two different current configurations.

Let, first, the current, flowing transverse to the external
magnetic field, in the direction {\em 1}, is homogeneously
distributed along the direction {\em 3}, i.e. along the external
magnetic field, $j(k_2,k_3)=\delta (k_3)\widetilde{j}(k_2)$. Then,
the magnetic induction produced by this current 
is parallel to direction {\em 3}, orthogonal to the current, and
parallel to the external magnetic field, and depends only upon the
 coordinate
$x_2$ across the external field:
 \bee\label{Ah1}
b_3(x_2)=\frac{\rmi}{(2\pi)^3}\int\frac{k_2\rme^{-\rmi k_2
x_2}\widetilde{j}(k_2) \rmd k_2}{k_2^2-\bar{\kappa}_3(0,0,
k_2^2)}=\frac{\rmi}{(2\pi)^3}\int\frac{k_2\rme^{-\rmi k_2
x_2}\widetilde{j}(k_2)\mu^{\rm pl}_{\rm tr} (k_2^2) \rmd
k_2}{k_2^2},\eend where \bee\label{mupllong} \mu^{\rm pl}_{\rm tr}
(k_\perp^2)=\left(1-\frac{\bar{\kappa}_3(0,0,k_\perp^2)}{k_\perp^2}\right)^{-1}.\eend
If the current, besides, is totally concentrated on the infinitely
thin surface coinciding with the coordinate plane ({\em 1,3}),
$\widetilde{j}(k_2)=(2\pi)^2 j$, where $j$ is a finite, constant
 linear current density, defined as the ratio of the
total current  to the length unit of the axis {\em 3}, its
magnetic induction far from the surface, behaves 
as\bee\label{planemag}b_3(x_2)\approx \frac {j\mu^{\rm pl}_{\rm
tr}(0)}{2}{\rm sgn} (x_2).\eend

Let, second, the current, flowing transverse to the external
magnetic field, in the direction {\em 1}, is homogeneously
distributed along the direction {\em 2}, orthogonal to the
external magnetic field, $j(k_2,k_3)=\delta
(k_2)\widetilde{j}(k_3)$. Then, the magnetic induction produced by
this current is parallel to direction {\em 2}, orthogonal to the
current and to the external magnetic field, and depends only upon
the coordinate $x_3$ along the external field:\bee\label{Ah2}
b_2(x_3)=\frac{-\rmi}{(2\pi)^3}\int\frac{k_3\rme^{-\rmi k_3
x_3}\widetilde{j}(k_3) \rmd k_3}{k_3^2-\bar{\kappa}_3(0,k_3^2,
0)}=\frac{-\rmi}{(2\pi)^3}\int\frac{k_3\rme^{-\rmi k_3
x_3}\widetilde{j}(k_3)\mu_{\rm tr}^{\rm pl} (k_3^2) \rmd
k_3}{k_3^2},\eend where \bee\label{mupltr} \mu_{\rm long}^{\rm pl}
(k_3^2)=\left(1-\frac{\bar{\kappa}_3(0,k_3^2,
0)}{k_3^2}\right)^{-1}.\eend If the current, besides, is totally
concentrated on the infinitely thin surface coinciding with the
coordinate plane ({\em 1,2}), $\widetilde{j}(k_3)=(2\pi)^2 j$,
where $j$ is a finite, constant current density per unit length
along the axis {\em 2}, its magnetic induction far from the
surface, behaves as \bee\label{planemag2}b_2(x_3)\approx \frac
{-j\mu_{\rm long}^{\rm pl} (0)}{2}{\rm sgn} (x_3).\eend
 According to
(\ref{kappa}), (\ref{mupltr}) \bee\label{mutr}\left(\mu^{\rm
w}_{\rm tr}(0)\right)^{-1}=\left(\mu^{\rm pl}_{\rm
long}(0)\right)^{-1} =1-\mathfrak{L}_\mathfrak{F}\geq 0\eend and
to (\ref{kappa}), (\ref{mupllong}) \bee\label{mulongpl}
\left(\mu^{\rm pl}_{\rm
tr}(0)\right)^{-1}=1-\mathfrak{L}_\mathfrak{F}+2\mathfrak{F}\mathfrak{L}_{\mathfrak{FF}}\geq
0 .\eend

 This work was supported by the Russian Foundation for Basic
Research (Project No. 05-02-17217) and the President of Russia
Programme (No. LSS-4401.2006.2), as well as by the Israel Science
Foundation of the Israel Academy of Sciences and Humanities.

\end{document}